\documentclass[sigconf]{acmart}
\usepackage[a-1b]{pdfx}
\usepackage[subtle]{savetrees}

\usepackage{booktabs} 
\usepackage[english]{babel}
\usepackage[export]{adjustbox}
\usepackage{mathrsfs}
\usepackage{csquotes}
\usepackage{multirow}
\usepackage{multicol}
\usepackage{array}
\usepackage{arydshln}
\usepackage{booktabs}
\usepackage{subfig}
\usepackage{siunitx}
\usepackage[inline]{enumitem}
\usepackage{xspace}
\usepackage{etoolbox}
\usepackage{wasysym}
\usepackage[skip=2pt]{caption}
\DeclareCaptionFont{xbf}{\bfseries\boldmath}
\usepackage{pbox}
\usepackage{siunitx}
\usepackage{makecell}

\makeatletter

\newcommand{\anon}[1]{\textcolor{black}{~(anon.)}}
\newcommand{\anoncite}[1]{\textcolor{black}{[anon.]}}

\newcommand{\flegend}{\vspace{-0.3cm}\subfloat{\includegraphics[trim=0 248 375 6, clip,width=0.7\columnwidth]{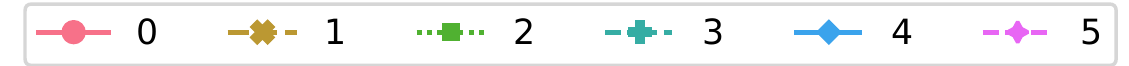}}}

\patchcmd{\maketitle}{\@copyrightspace}{}{}{}

 \makeatother
 
\newlist{inlinelist}{enumerate*}{1}
\setlist*[inlinelist,1]{%
  label=(\roman*),
}

\author{Mohammad Aliannejadi}
\affiliation{%
  \institution{University of Amsterdam}
}
\email{m.aliannejadi@uva.nl}

\author{Leif Azzopardi}
\affiliation{%
  \institution{University of Strathclyde}
}
\email{leif.azzopardi@strath.ac.uk}

\author{Hamed Zamani}
\affiliation{%
  \institution{University of Massachusetts Amherst}
}
\email{zamani@cs.umass.edu}

\author{Evangelos Kanoulas}
\affiliation{%
  \institution{University of Amsterdam}
}
\email{e.kanoulas@uva.nl}

\author{Paul Thomas}
\affiliation{%
  \institution{Microsoft}
}
\email{pathom@microsoft.com}

\author{Nick Craswell}
\affiliation{%
  \institution{Microsoft}
}
\email{nickcr@microsoft.com}

\copyrightyear{2021} 
\acmYear{2021} 
\setcopyright{acmcopyright}\acmConference[CIKM '21]{Proceedings of the 30th ACM International Conference on Information and Knowledge Management}{November 1--5, 2021}{Virtual Event, QLD, Australia}
\acmBooktitle{Proceedings of the 30th ACM International Conference on Information and Knowledge Management (CIKM '21), November 1--5, 2021, Virtual Event, QLD, Australia}
\acmPrice{15.00}
\acmDOI{10.1145/3459637.3482231}
\acmISBN{978-1-4503-8446-9/21/11}

\newcommand{\partitle}[1]{\vspace{1mm}\noindent\textbf{#1}}

\begin{document}

\fancyhead{}

\title[Analysing Mixed Initiatives and Search Strategies during Conversational Search]{Analysing Mixed Initiatives and Search Strategies\\ during Conversational Search}

\renewcommand{\shortauthors}{Aliannejadi, et al.}

\begin{abstract}
Information seeking conversations between users and Conversational Search Agents (CSAs) consist of multiple turns of interaction. While users initiate a search session, ideally a CSA should sometimes take the lead in the conversation by obtaining feedback from the user by offering query suggestions or asking for query clarifications i.e. mixed initiative. This creates the potential for more engaging conversational searches, but substantially increases the complexity of modelling and evaluating such scenarios due to the large interaction space coupled with the trade-offs between the costs and benefits of the different interactions. 
In this paper, we present a model for conversational search -- from which we instantiate different observed conversational search strategies, where the agent elicits: (i) Feedback-First, or (ii) Feedback-After. Using 49 TREC WebTrack Topics, we performed an analysis comparing how well these different strategies combine with different mixed initiative approaches: (i) Query Suggestions vs.~(ii) Query Clarifications. Our analysis reveals that there is no superior or dominant combination, instead it shows that query clarifications are better when asked first, while query suggestions are better when asked after presenting results. We also show that the best strategy and approach depends on the trade-offs between the relative costs between querying and giving feedback, the performance of the initial query, the number of assessments per query, and the total amount of gain required. While this work highlights the complexities and challenges involved in analyzing CSAs, it provides the foundations for evaluating conversational strategies and conversational search agents in batch/offline settings.
\vspace{-2mm}
\end{abstract}



\maketitle

\vspace{-2mm}
\section{Introduction} \label{sec:intro}
\textit{ Conversational Search }(CS) is an emerging area of research that aims to couch the information seeking process within a conversational format~\cite{DBLP:journals/dagstuhl-reports/AnandCJSS19,DBLP:journals/sigir/Culpepper0S18,croft2019interaction}.
 CS differs from the traditional query-response paradigm by providing more agency through improved query understanding and the persistence of the conversational context~\cite{DBLP:conf/sigir/YanSW16,DBLP:conf/chi/VtyurinaSAC17}. The exciting prospect of \textit{Conversational Search Agents} (CSAs) has spurred considerable research into the development of the underlying methods to support such agents. Of particular interest have been methods that facilitate mixed initiative approaches that aim to enhance the agent's understanding of the user's information need through query suggestions (i.e.~refinements, expansions, etc.) or through query clarifications (i.e.~questions that seek to clarify the query, elicit the user preferences, etc.)~\cite{DBLP:conf/acl/DaumeR18,DBLP:conf/www/ZamaniDCBL20,DBLP:conf/sigir/AliannejadiZCC19,DBLP:conf/ictir/KrasakisAVK20,DBLP:journals/corr/abs-2009-11352,aliannejadi21buidling}. This is because mixed initiative interactions are seen as a key property of a conversational search agent~\cite{DBLP:conf/chiir/RadlinskiC17} which has the potential to increase user engagement and user satisfaction~\cite{DBLP:conf/sigir/KieselBSAH18}. While various efforts have focused on building the infrastructure to support the inclusion of clarifying questions, and numerous methods proposed to generate or select good questions~\cite{DBLP:conf/chiir/BraslavskiSAD17,DBLP:conf/sigir/AliannejadiZCC19,DBLP:conf/www/ZamaniDCBL20,DBLP:journals/corr/abs-2006-10174,DBLP:conf/sigir/HashemiZC20,DBLP:conf/ecir/SekulicAC21,DBLP:journals/corr/abs-2103-06192}, little work has evaluated or compared the use of such methods within the context of a CS session in a batch/offline setting --- largely because the possible state space increases exponentially with interaction, coupled with the lack of a user model for CS. So while there has been considerable effort in the community to engage in single and mixed initiative conversations, little has been done to understand how they impact performance during CS sessions.

 In this work, our goal is to provide a user model for conversational search that can be used to evaluate mixed initiative approaches and conversational strategies. While asking query clarifications and offering query suggestions may lead to increases in user satisfaction in certain scenarios~\cite{DBLP:conf/sigir/KieselBSAH18}, it also imposes additional costs on the user; 
 the premise being that the investment in feedback will lead to greater returns later. 
 And so the costs and the expected gains associated with different mixed initiative approaches will determine whether eliciting or giving feedback is worthwhile compared to other actions that could be taken (e.g. re-querying or assessing)~\cite{azzopardi2018conceptual,Trippas2018}.
So how should an agent interact with a user? 
Should it ask a series of clarifications, and then present results, or present results, and then ask for clarifications? Or, not ask any clarifications? It is very much an open question what conversational search strategy should be employed in order to minimise the conversational cost while maximising the user's gain. And, how different mixed initiative approaches would influence the choice of strategy given the user's interactions (i.e. whether they assess more, give more feedback, or issue more queries). In this paper, we aim to provide insights into these research questions by modelling the CS process and then measuring the costs and benefits of different CS strategies and mixed initiative approaches.

\section{Background} \label{sec:rel}
Over the past few years, an increasing amount of attention has been directed toward developing methods that enable CS and the development of CSA, for example: ranking results given the conversation~\cite{DBLP:conf/sigir/YangQQGZCHC18,Dalton:2020:CAST}, generating clarifying questions~\cite{DBLP:conf/sigir/AliannejadiZCC19,DBLP:conf/www/ZamaniDCBL20,DBLP:conf/sigir/ZamaniMCLDBCD20}, studying system-initiative interactions~\cite{Wadhwa2021}, and presenting results~\cite{Spina:2017}. Less attention, however, has been focused on developing user models for evaluating CS which can be used to analyse CSAs and CS strategies.

One of the first CS systems was proposed by \citet{DBLP:journals/jasis/CroftT87}, called I$^3$R. It acted as an expert intermediary system, communicating with the user during a search session. Since then, other researchers have developed more elaborate approaches. For example, \citet{belkin1995cases} offered users choices in a search session using case-based reasoning. While, \citet{allen1999mixed} were among the first to study mixed initiative conversations, which they defined as ``\textit{a flexible interaction strategy in which each agent (human or computer) contributes what it is best suited at the most appropriate time}``. However, since then researchers have mainly focused on single-initiative interaction such as rule-based conversational systems~\cite{DBLP:conf/acl/WalkerPB01} and spoken language understanding approaches~\cite{225939,DBLP:journals/csl/HeY05}. Mixed initiative, though, provides a mechanism for the agent to improve its 
understanding of the user's information need by obtaining feedback by offering query suggestions (i.e. refinements to the query) or  query clarifications (i.e. questions that seek to clarify the query) ~\cite{DBLP:conf/acl/DaumeR18,DBLP:conf/sigir/AliannejadiZCC19,DBLP:conf/www/ZamaniDCBL20}. As previously mentioned, this idea of mixed initiative and the system taking agency has led to the development of CSAs. Inspired by models and work on conversations and dialogue systems (e.g. COR, etc.~\cite{belkin1995cases,McTear2002,Oddy1977,SITTER1992165}),   \citet{DBLP:conf/chiir/RadlinskiC17} developed a theoretical framework, that puts forward five key properties that a search system needs to have in order to be ``\textit{conversational}''. These properties are: 
\begin{itemize}[leftmargin=*]
\item \textbf{User Revealment} where the user discloses to the agent their information needs,
\item \textbf{Agent Revealment} where the agent reveals what the agent understands, what actions it can perform, and what options are available to the user,
\item \textbf{Set Retrieval} where the agent needs to be able to work with, manipulate and explain the sets of options/objects which are retrieved given the conversational context,
\item \textbf{Memory} where the agent tracks and manages the state of the conversation and the user's information need, and,
\item \textbf{Mixed Initiative} where both the agent and the user can take the initiative and direct the conversation search process.
\end{itemize}

\citet{azzopardi2018conceptual} extended this framework by defining the specific actions associated with these aspects. For example, within mixed initiative, they suggest that agents could seek to provide query suggestions or query clarifications that help to refine the user's information need, or seek to elicit the user's preferences, while users could, conversely, suggest refinements and disclose preferences. \citet{Trippas2018} examined how searchers interacted with intermediaries (who used the search engine), engaged actions and observed that searchers generally switched between \textit{query formulation} (user revealment) and result exploration (set retrieval), but also provided relevance feedback and clarifications. Following on from this work, \citet{Trippas2020} suggested a more general classification of the different interactions grounded by empirical studies -- their high level model delineates between: (i) \textit{discourse level actions}, that enable discourse management, grounding, visibility, and navigation and (ii) \textit{task level actions}, that are specific to the search such as handling queries, search assistance (e.g. clarifying queries), presenting results, and search progression. In another empirical study that analysed a number of CS datasets, \citet{vakulenko2019qrfa} found that users issue a query to the agent, and then the agent may respond with a request to clarify/refine the information need, or provide a list of results. The user could then respond by either issuing a new query, responding to the request, providing feedback, or assessing a result. They referred to this as the QFRA model~\cite{vakulenko2019qrfa}. They observed different patterns of behaviour such as \textit{query-feedback loops}, where several rounds of feedback to clarify/refine their query were observed, before results were assessed (\textit{Feedback-First}), and \textit{assessment-feedback loops}, where the user inspected results and then provided feedback to clarify/refine their query (\textit{Feedback-After}).
 
\citet{DBLP:conf/cikm/ZhangCA0C18} proposed a \textit{System Ask, User Respond} paradigm, which is akin to query-refinement, where after the initial request is made, the agent will ask for refinements/clarifications until it is confident enough to present results. ~\citet{Kaushik} presented a system sided workflow model consisting of the steps the agent takes when dealing with a user's request (e.g. handling greetings and error handling). Under their model when a query is entered, it is checked and if a clarification is needed, the user is asked for one clarification, else, the agent retrieves and presents three results. If these are not relevant, or the user wants to see more items, they can request another three results. Alternatively, they can request to view the document (or a summary of). Otherwise, they can issue a new query (or stop). A similar approach is presented in~\cite{Wambua2018} where up to eight rounds of feedback were performed. While ~\citet{Dubiel2020} presented a similar conversational workflow model, however, the agent asks for up to three rounds of feedback to refine the user's request, before presenting two results. More recently, \citet{lipani2021evaluating_css} they proposed a CS model based on exploring subtopics via a query-response paradigm, however, it did not consider feedback. In this work, we aim to explicitly model feedback and explore its impact within the conversational search process.

The various models proposed share a number of commonalities inherent to \textit{Interactive Information Retrieval} (IIR). Interaction consists of a number of turns based around: \textit{Query Formulation}, where the user expresses their query,  \textit{Result Exploration}, where the user examines results, and \textit{Query Reformulation}, where the user updates their query \cite{Marchionini1997,Sahib2012}. In terms of modelling, simulating, and evaluating the IIR process, the focus has largely been on considering sessions, rather than mixed initiatives. For example, the user browsing model which is at the heart of most IR evaluation metrics, assumes a user will pose a query, and then examine documents in a top down fashion~\cite{Carterette2011,Moffat2008}. For IIR, the model has been extended such that the user decides with some probability of examining the next document, or issuing a new query~\cite{Baskaya2012}, or examining a fixed number of documents before issuing a new query~\cite{DBLP:conf/sigir/Azzopardi11}. Through modelling the IIR process it has been shown trade-offs emerge between querying and assessing~\cite{DBLP:conf/sigir/Azzopardi11}, where, for example,  \citet{DBLP:conf/sigir/AzzopardiKB13} found that as query cost increased, users submitted fewer queries, and compensated by examining more results. But, to model CS, it is clear that the mixed initiative (feedback turn) needs to be explicitly considered within the user model. However, this will add additional complexities -- and invariably introduce new trade-offs because gathering feedback to refine or clarify the query will come at a cost -- which may or may not lead to more gain -- and so issuing a new query or assessing another result may be more beneficial. In~\cite{azzopardi2018conceptual,Trippas2020}, they point out that it is important for CSA to maximise the gain it delivers to the user, while trying to minimise the cost -- and thus maximise the rate of gain (following Grice's Maxims of Conversation~\cite{grice1975logic}). With the introduction of mixed initiative approaches and different CS strategies for engaging with a CSA there are many open questions. Is giving feedback (clarifications or suggestions) worth the cost, under what conditions is it beneficial, and, what type of CS strategy (feedback first or after) leads to a higher rate of gain? 

\begin{figure}
    \centering
    \vspace{-4mm}
    \includegraphics[width=8.5cm]{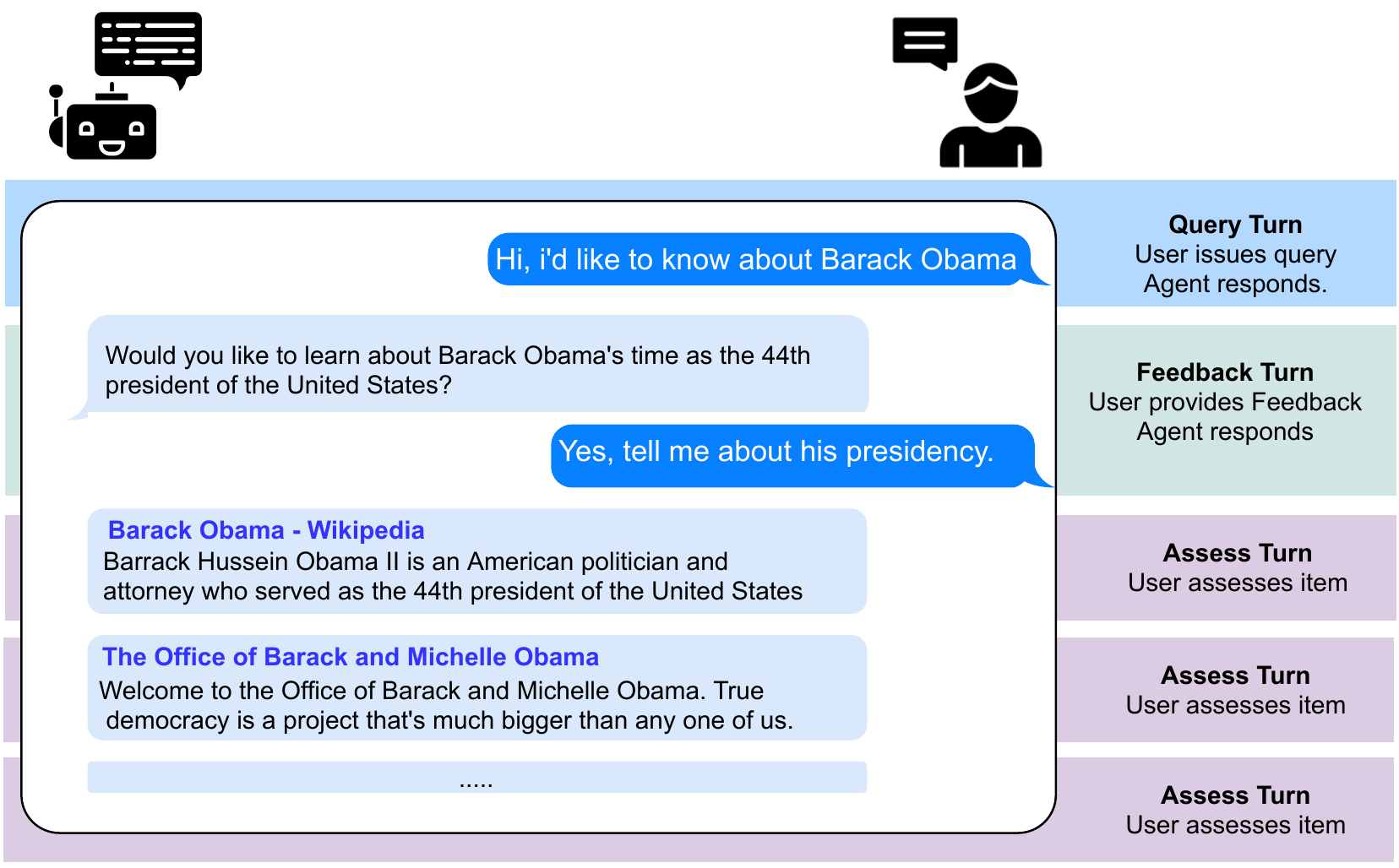}\\ 
    \includegraphics[width=8.5cm]{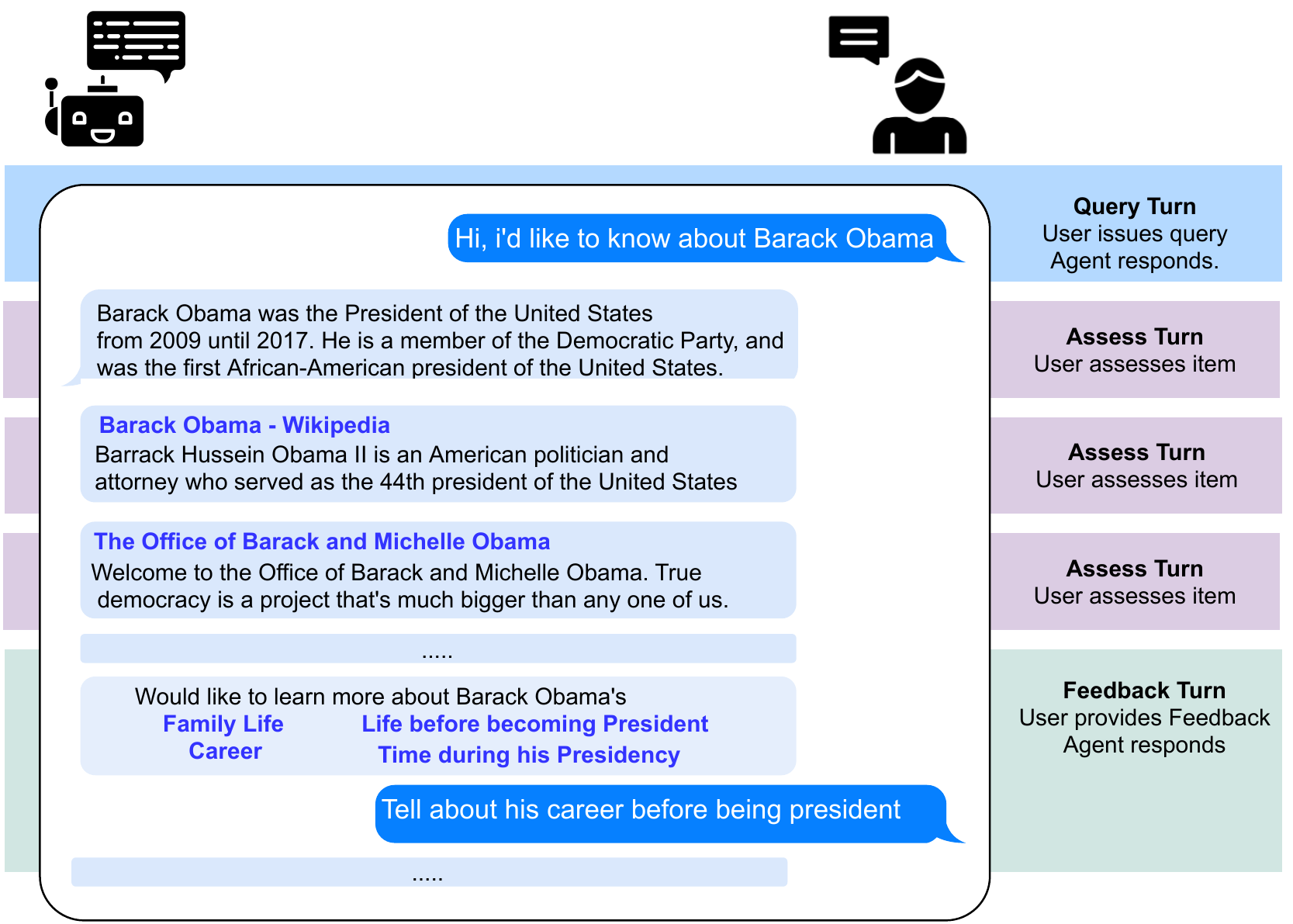} 
    \caption{Example chat-based CSA where one agent asks clarifications questions before presenting results (top), while the other agent asks query suggestions after presenting results (bottom). An open question is what strategy (e.g. initiate first or after results) and what type of mixed initiative (e.g. clarifications or refinements) leads to a better CS? 
     }
    \vspace{-6mm}
    \label{fig:csa_interface}
\end{figure}

\section{A Model of Conversational Search}
As previously discussed, conceptually conversational search can be seen as a special case of IIR~\cite{croft2019interaction} -- where the interaction between the user and the agent is based around conversational turns -- and the agent is more active in the seeking process through  mixed initiative interactions. So far we have not specified the details of the CSA, which could be: (i) a voice only CSA often via a virtual assistant ~\cite{kiseleva2016predicting,Trippas2020,Dubiel2020,DBLP:conf/chiir/TrippasSCJS18, Sahib2012}), (ii) a chat based CSA (that is in-situ within a platform like Slack~\cite{avula2017} or Telegram~\cite{Zamani2019}), (iii) an augmented search engine interface~\cite{Bosetti2017}, or (iv) a multi-modal virtual assistant~\cite{Zamani2019,johnston-etal-2014-mva}.  Given the wide range of CSAs, it is not possible to fully model them all -- so we need to make a number of assumptions about the type CSA we will model. Below, we outline what affordances the agent provides, and then what actions a user can take with respect to the search process e.g.~we are focusing on the salient search interactions (task level), and not on modelling error handling, chit-chat, etc.(discourse level). Given a CSA, we assume that the agent can initiate or respond as follows:
\begin{itemize}
\item[(i)] present results/answers to the user given their query, 
\item[(ii)] request additional feedback to update the query, 
\item[(iii)] or some combination of these. 
\end{itemize} 
The response (or combined response, and order of) will depend on numerous factors e.g.~the modality of the agent, bandwidth available, and agent capabilities.  We assume that the user in turn can perform the following actions during the search process:

\begin{itemize}
    \item[(i)] issue a query, 
    \item[(ii)] provide feedback, or
    \item[(iii)] assess a result/answer.
\end{itemize}
 
Given the affordances of the agent and how the user can interact with the agent, then we can conceptualise the conversational search process as a series of turns consisting of three main turn types: ($\tau_Q$) query turn, ($\tau_F$) feedback turn, and ($\tau_A$) assessment turn. 

As previously mentioned, how the agent responds will be different depending on the type of agent, its modality, and the current context. For example, (i) a voice only CSA needs to be very sensitive to the limited and serial bandwidth of speech, and so responses are likely to be shorter, (ii) a multi-modal CSA could present a more detailed combined response i.e.~a search engine result page that asks a number of requests for feedback (via query suggestions, facets, etc.), provides many results, etc., while (iii) a chat based CSA agent  has some restrictions on screen space and bandwidth within the chat window, but has the advantage over voice only CSAs because the conversation is persistent so the user can refer back to options, etc.. For the purposes of this work, we will assume that the CSA is a chat-based CSA like that in Fig.~\ref{fig:csa_interface} which represents the interfaces explored in~\cite{avula2017,Zamani2019,Kaushik}. Of note is that depending on the interface and its modality, the cost of different conversational turns will vary -- and this will impact on how much gain the user accumulates from their conversation with the agent. As proposed in~\cite{azzopardi2018conceptual,DBLP:conf/chiir/TrippasSCJS18}, we also assume that a user wants to maximise the amount of gain they receive from the system, while trying to minimise the cost of the conversation (where the cost could be the total number of turns, or the total time taken to perform those turns). Note that the assumed objective does not necessarily mean that a user prefers a shorter conversation, but one that yields a higher rate of gain. 
An open question then is how should a user and CSA work together in order to maximise the user's rate of gain? Should the user give many rounds of feedback, or should they examine some results, and then give feedback to the system? Should the CSA request more feedback, or provide more results? 
\vspace{-2mm}
\subsection{Modelling the Conversation Search Process}
To model the conversation search process, we draw upon the previous work that has conceptualised conversational and interactive search, with the aim to formalise the key actions/turns described above within a \textit{Markov Decision Process} (MDP) model (as done in ~\cite{Baskaya2012,maxwell2016agents,Moffat2008,DBLP:conf/sigir/AzzopardiTM19,thomas2014modeling} for IIR).
In these works, MDPs  (or variants of) have been used to represent key decisions that users make when searching and interacting with a system and their key actions that they take. For example, the simple \textit{User Browsing Model} (UBM) that underpins most metrics in IR~\cite{Carterette2011}, assumes that a user will issue a query, then assess a result item accumulating some gain if it is relevant. The user then decides to either continue and assess the next item with some probability of continuing or stop examining result items~\cite{Moffat2008,DBLP:conf/cikm/MoffatTS13}. 
In~\cite{thomas2014modeling,maxwell2016agents}, the UBM was extended to IIR to include additional decision points to model session search. However, because conversational search affords mixed initiative where feedback can be elicited from the user, the process is much more complicated. This additional affordance means that we need to include other decision points to capture these conversational turns -- and integrate the feedback process with the browsing process within the larger context of the search session.

\begin{figure}
    \centering
    \vspace{0mm}
    \includegraphics[width=8cm]{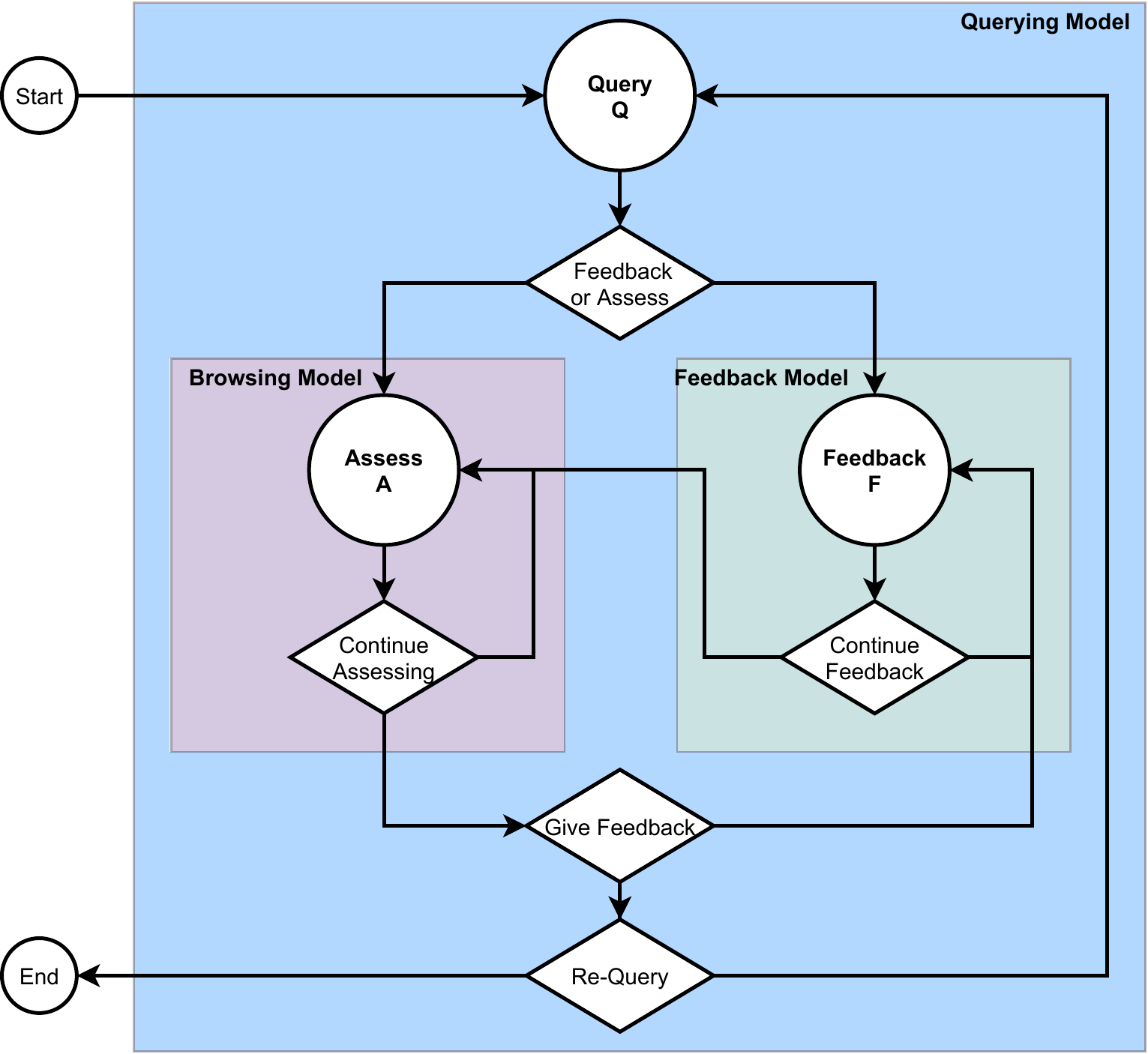} 
    \caption{The User Model of Conversational Search which is composed of three sub-components the Querying, Browsing and Feedback Models. Diamonds represent user decision points, while circles represent the action/turn taken. }
    \vspace{-2mm}
    \label{fig:markov_model}
\end{figure}

Fig.~\ref{fig:markov_model} presents our model of the CS process -- where we have broken up each of the user choices into binary decisions (denoted by diamonds), and the three actions/turns with circles. We assume that the user starts the search process by issuing a query ($\tau_Q$). Given the query, the agent responds either with a list of results, clarifications, suggestions, or a combination of. Essentially the agent presents the ``\textit{search engine result page}'' either via a web page or a chat bot via text (as in the Figure~\ref{fig:csa_interface}) or through speech. Given the response, the user may decide to either inspect results or give feedback depending on what options are presented by the agent. If they choose to assess a result ($\tau_A$) they follow the typical UBM, shown in light purple. While if they choose to provide feedback ($\tau_F$) then they follow the User Feedback Model (UFM) shown in light green.

Following the UBM, if a user performs a $\tau_A$ turn, then they inspect a result item, where it is assumed that they will accumulate some gain if the result is relevant, and then they need to decide whether to perform another $\tau_A$ turn, or not~\cite{Moffat2008,DBLP:conf/cikm/MoffatTS13}. The decision to continue, would of course, depend on a number of factors such as how much gain has been accumulated, how many items have been examined/assessed, etc.~\cite{DBLP:conf/cikm/MoffatTS13}.  Once they decide to stop assessing, the user may decide to give feedback to the agent, in order to refine/expand their current query, or not. If not, the user can then decide whether to re-formulate their query, in which case repeating the process. Otherwise, they stop searching. Similarly, in the case, where the user decides to give feedback ($\tau_F$), the can provide feedback to the agent, where it is assumed that the agent will provide an updated response, and then the user needs to decide whether to perform another round of feedback ($\tau_F$), or not. Once they stop giving feedback, the user can decide whether to go back to assessing, or not. And, if not they then need to decide whether to re-query, or stop searching altogether.

Fig.~\ref{fig:csa_interface} presents two example conversations. In the top example, the user issues a query, and the agent responds by asking a query clarification, the user responds and then the agent presents a number of results. In the bottom example, the user issues a query, and the agent responds with a number of results, followed by a request for feedback via query suggestions, to which the user responds, and the conversational continues.  However, the space of possible sequences of different turns grows rapidly. And, herein lies the complexity of evaluating conversational search -- after $t$ turns, the number of possible conversational sequences, is approximately $3^{t}$ for a fully mixed initiative CSA. Nonetheless, the number of possible sequences of conversational turns exponentially increases with the number of turns. This presents an open challenge in evaluating Conversational Search Agents.

\subsection{Instantiating the User Model for CS}
In order to make the problem tractable, we need to reduce the number of possibilities so that we can simulate and then evaluate the CS process. Grounded by observed behaviours from~\cite{vakulenko2019qrfa}, we propose two strategies for conversational search: 

\begin{itemize}
     \item \textbf{Feedback First (FF)}: where the user performs a query-feedback loop before assessing. That is, after querying the user given $F$ rounds of feedback, before assessing $A$ items. 
     \item \textbf{Feedback After (FA)}: where the user performs assessment-feedback loops, where after  assessing $A$ items, the user gives feedback, and then repeats the process $F$ times.  
\end{itemize}
    
These two interaction models represent two ``\textit{pure}'' strategies that users/agents might evolve/apply. The first approach, Feedback-First, represents a CSA that is like a Librarian or Booking agent.
Here, the agent asks the user a number of clarifying questions or makes a number of suggestions to refine the user's information need before presenting results to the user. The second approach, of Feedback After, represents a more exploratory search setting where the user learns about the topic, and then provides feedback to the agent to progress their search through the topic space. While in practice it is likely that the optimal CS strategy would be a mixture of FA and FF, investigating these strategies is feasible, and has not been previously evaluated. Given these two  strategies, we aim to draw insights into how and when they are more successful and under what conditions. 
For example, how do the performance of the initial query, the cost of turns, the type of feedback, and the searcher's strategy interact and influence performance?

\vspace{-3mm}
\subsection{Evaluating the Gain and Cost of CS}
\label{sec:gaincost}
While evaluation in a traditional IR setting is primarily concerned with measuring the expected utility of a ranked list, CS introduces an interaction space that grows exponentially with interaction. This makes evaluating different strategies and methods more complicated, because the different interactions have different costs and provide different benefits. For example, giving feedback comes at a cost, on the hope that it will lead to accruing more gain later on. Obviously, if feedback turns are expensive, and they don't lead to greater increased gain, then the ``conversational'' part of the search may not be beneficial. To represent the costs associated with each action, we model the cost for the conversational turns $\tau_Q$, $\tau_F$ and $\tau_A$ as $c(\tau_Q$), $c(\tau_F$) and $c(\tau_A)$, respectively. The cost associated with each turn will depend on the response and the modality of the CSA. Consequently, when considering which CS strategy or which CSA is better than another, we cannot be agnostic to the cost of the conversation. Both cost and gain arising from the CS need to be measured.
In the CS setting, we can generalise the cumulative gain metric from traditional IR evaluations~\cite{jarvelin2017ir} to turns, where the \textit{Turn-based Cumulative Gain} of a sequence of conversational turns is: $G(t_1,\dots,t_T) = \sum_{i=1}^{i=T} g(t_i) \label{eq:gain}$
where $T$ is the total number of turns, and $g(t_i)$ is the gain obtained from the $i$th turn, and $t_i$ is either $\tau_Q$, $\tau_F$ or $\tau_A$.
As each turn comes at a cost, then the total cost, is: $C(t_1,\dots,t_T) = \sum_{i=1}^{i=T} c(t_i)\label{eq:cost}$,    
where $c(t_i)$ is the cost of performing the $i$th turn. The subsequent rate of gain, can then be calculated as total gain divided by the total cost:
$R(t_1,\dots,t_T) = \frac{G(t_1,\dots,t_T)}{C(t_1,\dots,t_T)}\label{eq:rate}$     
While discounting or session based metrics could be applied \cite{kalervo2008sdcg,azzopardi2017ift,lipani2021evaluating_css}, we leave such directions for further work, as it is not clear how the discounts would or should be applied in this context.
\vspace{-2mm}
\section{Research Questions}
In the context of conversational information seeking, where a user wants to explore a topic, and find out about various facets of the topic, through a conversational chat bot interface (like the one in Fig.~\ref{fig:csa_interface}), we aim to obtain insights into the following research questions:
\begin{itemize}[leftmargin=*,nosep]
    \item How does the conversational strategy (Feedback-First or Feedback-After) affect  performance?
    \item How does the mixed initiative approach (Clarification or Suggestion) affect the performance?
    \item How is the strategy and/or approach affected by the quality of the initial query?
    \item How is the strategy and/or approach affected by changes in the cost of turns?
\end{itemize}

\section{Experimental Method} \label{sec:exp}
To answer our research questions, we have opted to undertake a simulated analysis as done in previous works on IIR~\cite{DBLP:conf/jcdl/JordanWG06,DBLP:conf/sigir/AzzopardiRB07,DBLP:journals/ir/KeskustaloJP08,DBLP:conf/clef/HuurninkHRB10,Maxwell2015,maxwell2016agents,Zhang2020}. This is because the space of possible interaction sequences is very large and evaluating the different combinations would not be feasible in a user study. However, we do ground our analysis by conducting a user study to obtain estimates of the costs of performing different turns using a text based CSA (as in Fig.~\ref{fig:csa_interface}).

\partitle{Collection.} Following \citet{DBLP:conf/sigir/AliannejadiZCC19} we use the topics created as part of the TREC Web Track from 2009 to 2012, based on the ClueWeb09-Category B collection. The collection consists of 198 topics. Each topic consists of a series of facets that the user would like to explore -- making them suitable to explore in a conversational manner because clarifications and suggestions can help refine or redirect the search towards the different facets that the user wants to explore.
To ensure having a reasonable space of exploration, we filter out the topics that have fewer than four facets or fewer than ten relevant documents. These steps lead us to 49 topics with a total of 211 facets (approx 4.3 per topic).

\partitle{Conversational Search Agent.}
For our study, the CSA is defined by: (i) the conversational strategy that it employs either \textit{Feedback-First} (FF) or \textit{Feedback-After} (FA), (ii) the mixed initiative approach of \textit{Query Clarification} (QC) or \textit{Query Suggestion} (QS), (iii) the number of rounds of feedback that it offers (F), and (iv) the number of result items it presents to be assessed by the user (A). 

\partitle{Retrieval of Results.}
Given the query, and any subsequent clarifications or suggestions, we pre-process the query terms (i.e.~stopword removal and stemming) and submit it to the retrieval system.
To retrieve the ranked list of documents, we use an extension of the \textit{Query Likelihood Model} (QLM) for CS proposed in \citet{DBLP:conf/sigir/AliannejadiZCC19} with the suggested parameters. The model is a linear interpolation of the language model based on the query submitted by the user, and the language model based on the feedback.
Once the results are retrieved, the result lists are filtered, and only previously unseen result items are presented to the user. We assume the CSA has a memory of what results the user has already seen. 

\partitle{User Interactions.}
Following the user model presented in Fig.~\ref{fig:markov_model}, we assume that the user follows the search strategy given by the specific CSA. Below we describe how our simulated users generate queries which they issue during query turns, and then describe the feedback presented to them during feedback turns.

\partitle{Query Generation (Q).}
To generate the queries we employed the approach given by~\cite{DBLP:conf/jcdl/JordanWG06,DBLP:conf/sigir/AzzopardiRB07}. For each topic, a language model is created given the set of documents relevant to the topic. Then, to generate a query of length $L$, terms are sampled without replacement from the top 20 terms given their relative entropy in the language model.

\partitle{Feedback \textbf{(F)} - Query Clarifications and Query Suggestions.} 
Given the query issued, we assume that the agent is able to either (i) ask clarifying questions or (ii) provide query suggestions. The answers to the clarifying questions, or selection of the query suggestions, are then used to improve the query representation.
Each ${\tau_F}$ turn is expected to lead to a better query representation, which in turn should lead to improved query performance.
We take two approaches for simulating feedback:
\begin{itemize}[leftmargin=*]
    \item \textbf{Query Clarifications.} 
        For clarifications, we used the query clarifications from the Qulac dataset~\cite{DBLP:conf/sigir/AliannejadiZCC19} along with the human responses. We followed \cite{DBLP:conf/ictir/KrasakisAVK20} and pre-processed the data to remove redundant clarifications and low quality answers.
    \item \textbf{Query Suggestions.}
        For suggestions, we used the same query generation algorithm as before to generate additional terms used as suggestions.
     As shown in Fig.~\ref{fig:csa_interface}, the user is presented with four query suggestions, and when giving feedback the user selects a suggestion at random.
        \end{itemize}
Each successive round of feedback given, adds additional terms to the original query. We checked the performance of the resulting expanded queries given the query clarifications or suggestions and found that there was no significant difference between the two approaches at neither P@10 nor P@20. 

\partitle{Calculating the Gain.}
To calculate the gain, we follow Section~\ref{sec:gaincost}, where we assume that the user only accumulates gain on an assessment turn ($\tau_A$), where $g(\tau_A)=1$ when the user assesses a previously unseen relevant item, otherwise $g(\tau_A)=0$ and for the other turns: $g(\tau_Q)=g(\tau_F)=0$. That is, a user only received gain when they are provided with relevant and novel information during the conversation (as done in~\cite{DBLP:conf/sigir/Azzopardi11,Baskaya2012,DBLP:conf/sigir/SmuckerC12,maxwell2016agents}). 

\partitle{Estimating the Cost.}
To ground the estimation of the costs for each of the conversational turns, we conducted a user study where we designed four crowdsourcing tasks (HITs) on Amazon Mechanical Turk\footnote{\url{http://mturk.com}}. In all of our tasks, we first showed the user a search topic description (from a total of five search topics from the TREC Web Track). Our choice of topic was based on their difficulty and type (informational and faceted), aiming to cover a wide spectrum of search tasks in the study. Each search session started with a query from the user. Once the user clicks the Search button, they were shown either a result snippet or document and asked to judge its relevance (definitely relevant, possibly relevant, non-relevant). As soon as the worker assessed one snippet (or document), we show the next snippet (or document). We repeated this process for five results. After assessing the fifth result, we instructed the workers to either: (i) reformulate their query to look for a different facet of the topic, (ii) provide feedback by answering a clarifying question, or (iii) select one of the four query suggestions.
Each HIT provided data for 20 result assessments, 1 or 4 queries, and 3 rounds of feedback. We had 81 workers undertake the HITs, who submitted 144 queries, assessed 1,280 result snippets, 1000 result web pages, and provided 268 responses to feedback. The average time taken to issue a query was $29.3$ seconds, to assess a result snippet was $6.3$ seconds, to assess a result web page was $17.0$ seconds, while the average time to provide feedback was $8.3$ seconds. While in practice the cost of selecting suggestions vs.~providing clarifications will differ depending on the implementation we wanted to compare the two approaches as fairly as possible -- and thus kept the feedback costs the same between mixed initiatives.

To calculate the total cost, we follow Section~\ref{sec:gaincost} where we set the costs as: $c(\tau_Q)=29.3$, and $c(\tau_F)=8.3$. For estimating $c(\tau_A)$ we draw upon past work~\cite{DBLP:conf/sigir/SmuckerC12,Baskaya2012}, where the cost of assessing an item depends on its relevance, such that: $c(\tau_A) = c(s)+c(d)\big(P(C=1|R=1) + P(C=1|R=0)\big)$  where the cost of inspecting a snippet is $c(s)=6.3$, the cost of inspecting the document is $c(d) = 17.0$  and $P(C=1|R)$ is the probability of clicking on the item given its relevance. In this work, we set  $P(C=1|R=1)=1$ and $P(C=1|R=0)=0$, and thus assume the user only inspects relevant items but always pays the cost of examining the snippet regardless of relevance. One could explore more sophisticated click models, e.g., to account for position and trust bias. We leave exploration of these options for future work.   While this is a very optimistic setting -- we found including mis-clicks on non-relevant items or lower probabilities of clicking relevant items had little impact on which strategy/approach resulted in a higher rate of gain -- only that changes lowered the overall rate of gain of all conditions. Instead, our findings show that the relative costs between querying and giving feedback play a much larger role in the choice of strategy/approach (see \textsection\ref{sec:cost-changes}).
We leave modelling other variations in cost for future work.

\partitle{Simulated Analysis.}
To perform the analysis, we first decided on the CS strategy (i.e.~FF or FA) and a mixed initiative approach (i.e.~QC or QS) the agent would adopt, and then simulated the interaction as follows. For each topic, we assume that a user submits a query to the agent. The user either gives feedback and then examines results, or examines results and then gives feedback depending on the CS strategy. We recorded costs and gains as the number of queries (Q) is varied from 1 to 15, the number of rounds of feedback (F) is varied from 1 to 10, and the number of results assessed (A) is varied from 1 to 20. The total number of conversational turns for the FF strategy is $Q \times (F+A)$ and for FA strategy is $Q \times (F+1) \times A$.  The entire process was repeated 20 times for each of the 49 topics, for each strategy and mixed initiative (2x2). To explore the influence of query quality on CS we varied the length of queries during the generation process from 1 to 4. This resulted in over 12 million simulated CS sessions being generated for our analysis\footnote{Code and data: \url{https://github.com/i2lab/cikm21-conversational-search-strategies}.}.

\section{Results and Analysis}
To focus the presentation of our results, we will constrain our reports to the interactions within ten minutes (of simulated time) -- and unless stated otherwise, present the results when the starting query is of length two (L=2).

\subsection{Conversational Trade-offs}
\begin{figure}
    \vspace{-.5cm}
    \subfloat[FF-QC]{\includegraphics[height=3.3cm]{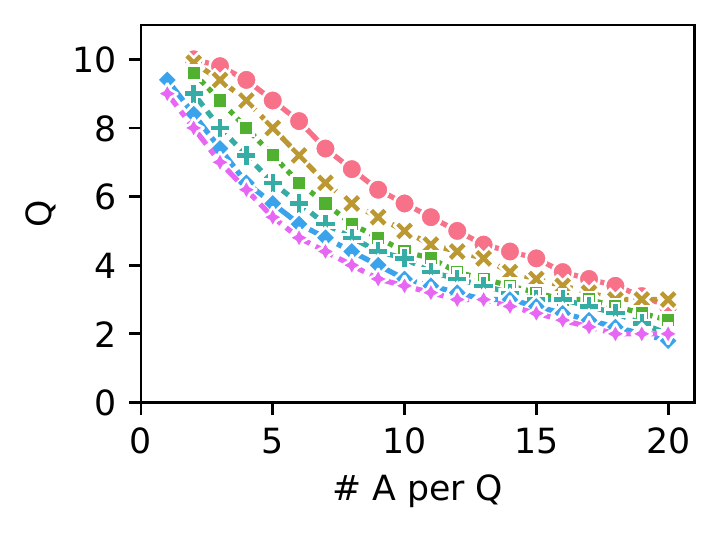}}
    \subfloat[FA-QC]{\includegraphics[height=3.3cm,trim=20 0 0 0, clip]{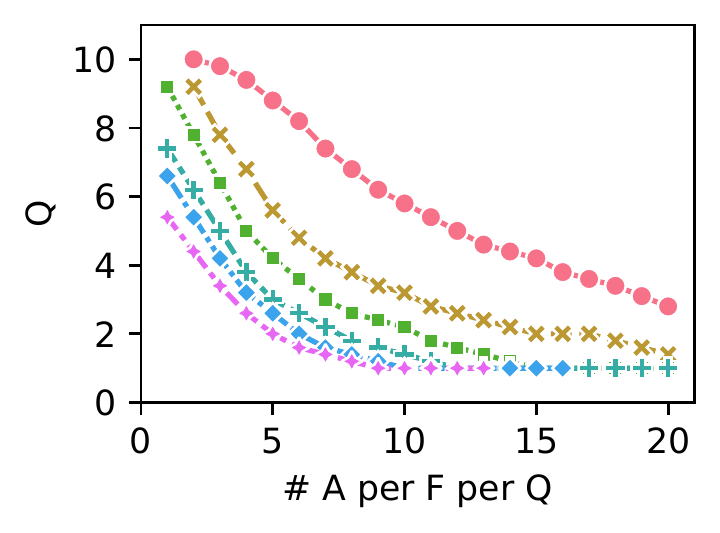}}
    \vspace{-3mm}
    \subfloat[FF-QS]{\includegraphics[height=3.3cm]{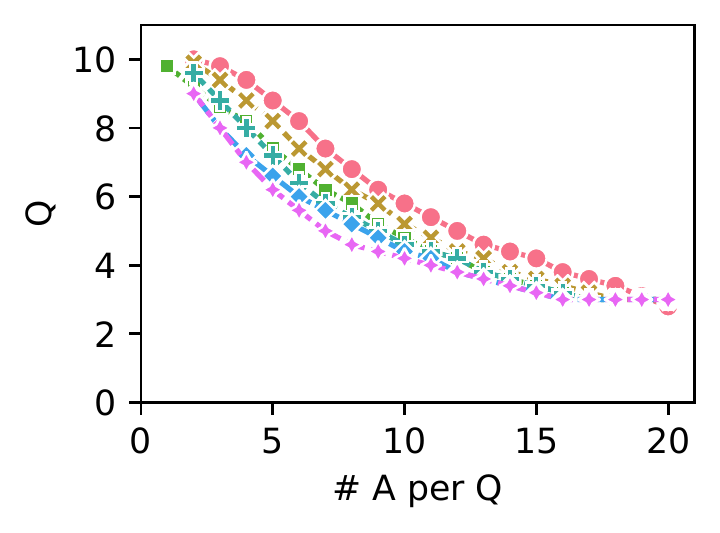}}
    \subfloat[FA-QS]{\includegraphics[height=3.3cm,trim=20 0 0 0, clip]{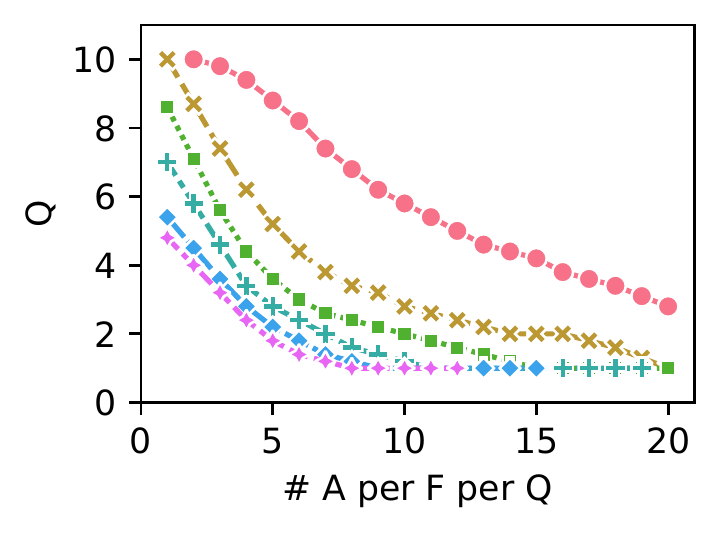}}
    \vspace{-3mm}
    \flegend
    \vspace{4mm}
    \caption{The plots show the trade-off between querying (Q) and assessing (A) for different levels of feedback (F) for the two strategies and two mixed initiatives. Left: Feedback First and Right: Feedback After. Top:  Query Clarification and Bot.: Query Suggestion. For clarity, only F$\le$5 and Q$\le$10 is shown.}
    \label{fig:trade-offs}
    \vspace{-4mm}
\end{figure}

To provide some insights into the trade-off between the different conversational turns, in Fig.~\ref{fig:trade-offs} we have plotted queries (Q) vs assessments (A) for the different rounds of feedback (F) for each conversational strategy and mixed initiative approach. The plots show the interactions for approx.~10 minutes of simulated time.  From the plots we can see that as more rounds of feedback are included, the number of queries and the number of assessments decrease -- because given a conversational search session of a similar length, taking an F turn comes at the expense of taking an alternative turn.  When we compare the Feedback-First strategy (left) to a Feedback-After strategy  (right), we can see that Q and A decrease by a greater amount. Recall, though, the subtle difference between conditions: in the FF strategy users perform F rounds of feedback, then examine A items, while in the FA strategy every round of feedback means they examine A items.  If we examine the query plots, we can see that as the number of rounds of feedback (F) increases, then the number of queries issued (Q) and the number of assessments performed (A) decreases. For the FA strategy the number of possible queries decreases at a much faster rate as F increases because of the successive rounds of assessing after each round of feedback. Given the space of possible conversational sequences, we  now turn our attention to comparing how well the different combinations of strategy and initiative perform. To make our comparisons we will be reporting the rate of gain, because different combinations lead to different session lengths, depending on the conversation turns taken and the relevant items found -- also reporting the rate of gain also means we can visualize the performance w.r.t the different number of interactions.

\subsection{Conversational Strategy vs.~Mixed Initiative}
To answer our main research question of how performance is affected by the conversational strategies: Feedback-First (FF) or Feedback-After (FA), vs. the Mixed Initiative (MI) approaches: Query Clarification (QC) or Query Suggestion (QS), we considered how the rate of gain (R) changed as the number of assessments (A) and levels of feedback (F) were varied for different CSA combinations. 

\begin{figure}[h]
\vspace{-6mm}
    \subfloat[FF-QC]{\includegraphics[height=3.45cm]{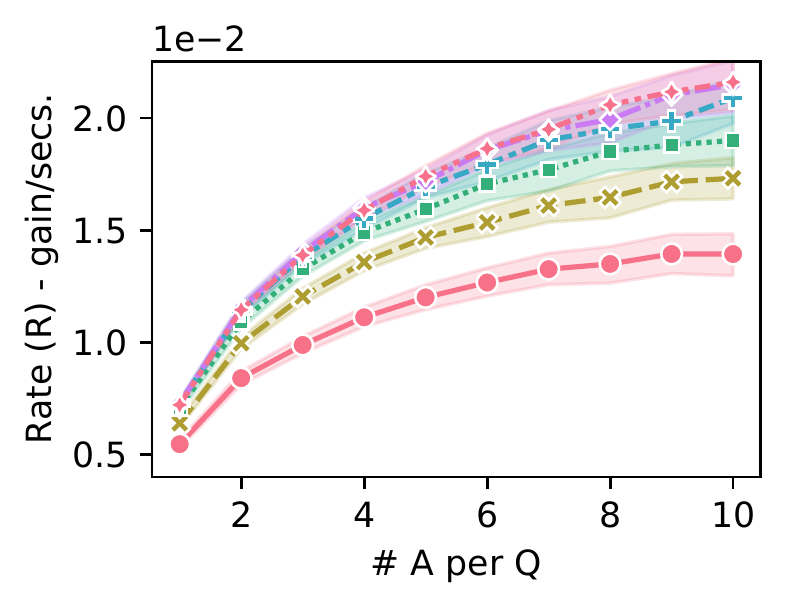}}
    \subfloat[FA-QC]{\includegraphics[height=3.45cm,trim=20 0 0 0, clip]{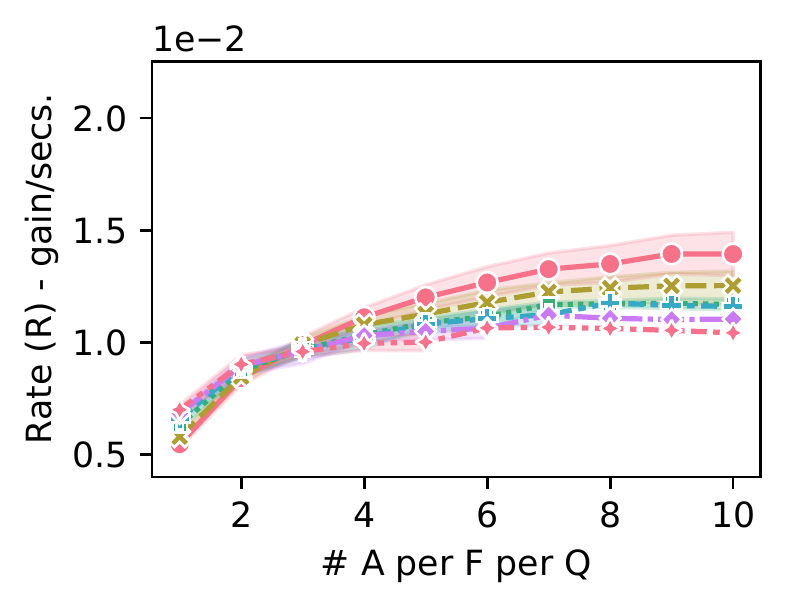}}
\vspace{-3mm}

    \subfloat[FF-QS]{\includegraphics[height=3.45cm]{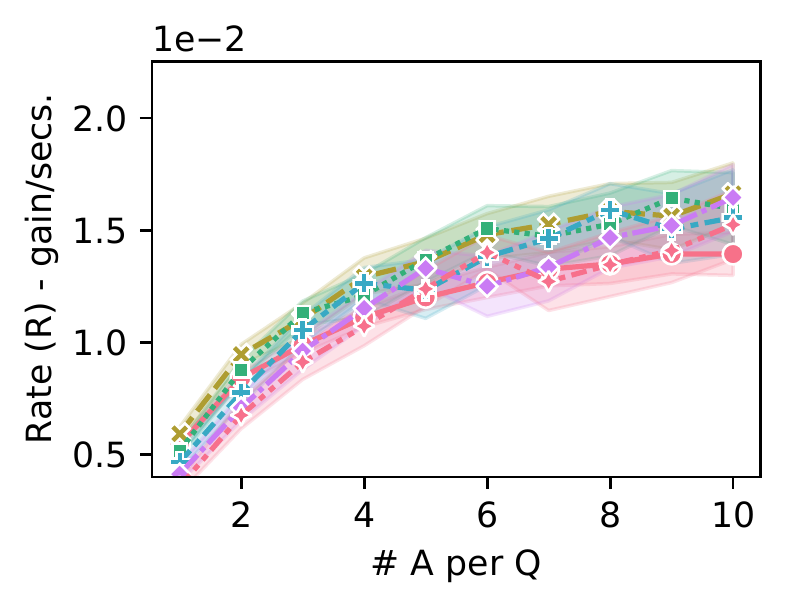}}
    \subfloat[FA-QS]{\includegraphics[height=3.45cm,trim=20 0 0 0, clip]{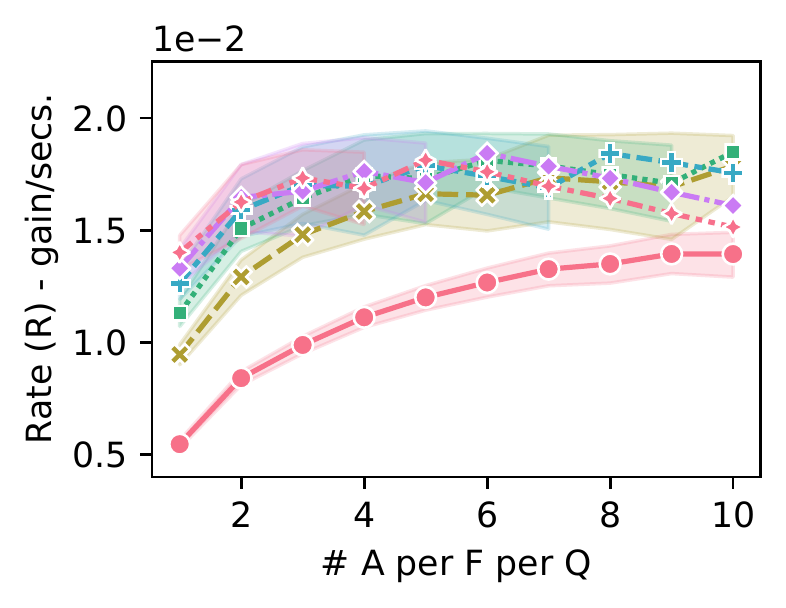}}
    
    \flegend
    \caption{The Rate of Gain (R) by the number of Assessments (A) for  different levels of feedback (F). Top: Query Clarification, Bot: Query Suggestion, Left: Feedback First, and Right: Feedback After. For clarity, only F$\le$5 and Q$\le$10 is shown.}
    \label{fig:cs-mi}
\end{figure}

First, we can how the different MI approaches perform for the FF strategy by inspected the left hand plots in Fig.~\ref{fig:cs-mi}. The top left plot shows that when query clarifications are employed it leads to substantial increases in the rate of gain over the baseline (i.e. when no feedback is given/provided $F=0$). Additional rounds of feedback increase the rate of gain but with diminishing returns. While  as the number of assessments per query (A) increases, the rate of gain also increases. This makes sense, because the investment in improving the query means that more relevant information is surfaced later on. However, when query suggestions are offered with the FF strategy, we observed a similar trend in the bottom left plot, but not as pronounced. In fact, after two rounds of query suggestions the rate of gain starts to decrease such that five iterations results in similar gain to the no feedback baseline.

\begin{table}[t]
    \centering
    \vspace{-2mm}
    \caption{For each combination of CSS x MI and for the no feedback condition (F$=0$), the best settings (query Q$^*$, assessment A$^*$, feedback F$^*$, cost C$^*$, and rate of gain R$^*$), on average, to achieve a gain (G) of 1, 5 and 9 when the starting query is length 2.}
    \begin{tabular}{lllllll}
        \toprule
            CSSxMI & G & Q$^*$ & A$^*$ & F$^*$ & C$^*$ & R$^*$ \\
        \midrule
            No Feedback &  1 & 1 &   7 &  0 &    96 &  0.014 \\
                  FF-QC &  1 & 1 &   5 &  1 &    89 &  0.015 \\
                  FA-QC &  1 & 1 &   5 &  1 &   123 &  0.012 \\
                  FF-QS &  1 & 1 &   5 &  1 &    90 &  \textbf{0.016} \\
                  FA-QS &  1 & 1 &   2 &  1 &    80 &  0.015 \\
            \midrule
              No Feedback & 5 & 3 &   9 &  0 &  338 &  0.014 \\
                    FF-QC & 5 & 1 &  11 &  5 &  209 &  \textbf{0.023} \\
                    FA-QC & 5 & 2 &  10 &  1 &  409 &  0.013 \\
                    FF-QS & 5 & 2 &  10 &  1 &  278 &  0.017 \\
                    FA-QS & 5 & 1 &   5 &  3 &  256 &  0.019 \\
            \midrule
             No Feedback &  9 & 7 &  10 &  0 &  791 &  0.011 \\
                   FF-QC &  9 & 2 &  11 &  5 &  407 &  \textbf{0.022} \\
                   FA-QC &  9 & 4 &  11 &  1 &  842 &  0.011 \\
                   FF-QS &  9 & 5 &  10 &  1 &  635 &  0.014 \\
                   FA-QS &  9 & 1 &   9 &  5 &  541 &  0.016 \\
        \bottomrule
        \end{tabular}
    \label{tab:best_r}
    \vspace{-5mm}
\end{table}

The plots on the right hand side of Fig.~\ref{fig:cs-mi}, show how the two mixed initiatives perform under the FA strategy. When query clarifications are offered after the user assesses items, then the rate of gain initially is higher than the baseline until A increases past three result items, then the strategy becomes less effective and the rate of gain drops below baseline (top-right plot). Interestingly, for query suggestions we see that rate of gain is much higher when suggestions are taken afterwards, and it is only if the user assesses more items per round of feedback does the rate of gain start to decrease and tend towards the baseline (bottom-right plot). Here, we see the query suggestions improve the initial query and bring more relevant information back in subsequent assessment turns -- but crucially assessing only a few items and then providing feedback leads to the highest rate of gain for this mixed initiative approach.

\begin{figure}[t]
\vspace{-3mm}
    \centering
    \includegraphics[width=0.8\columnwidth]{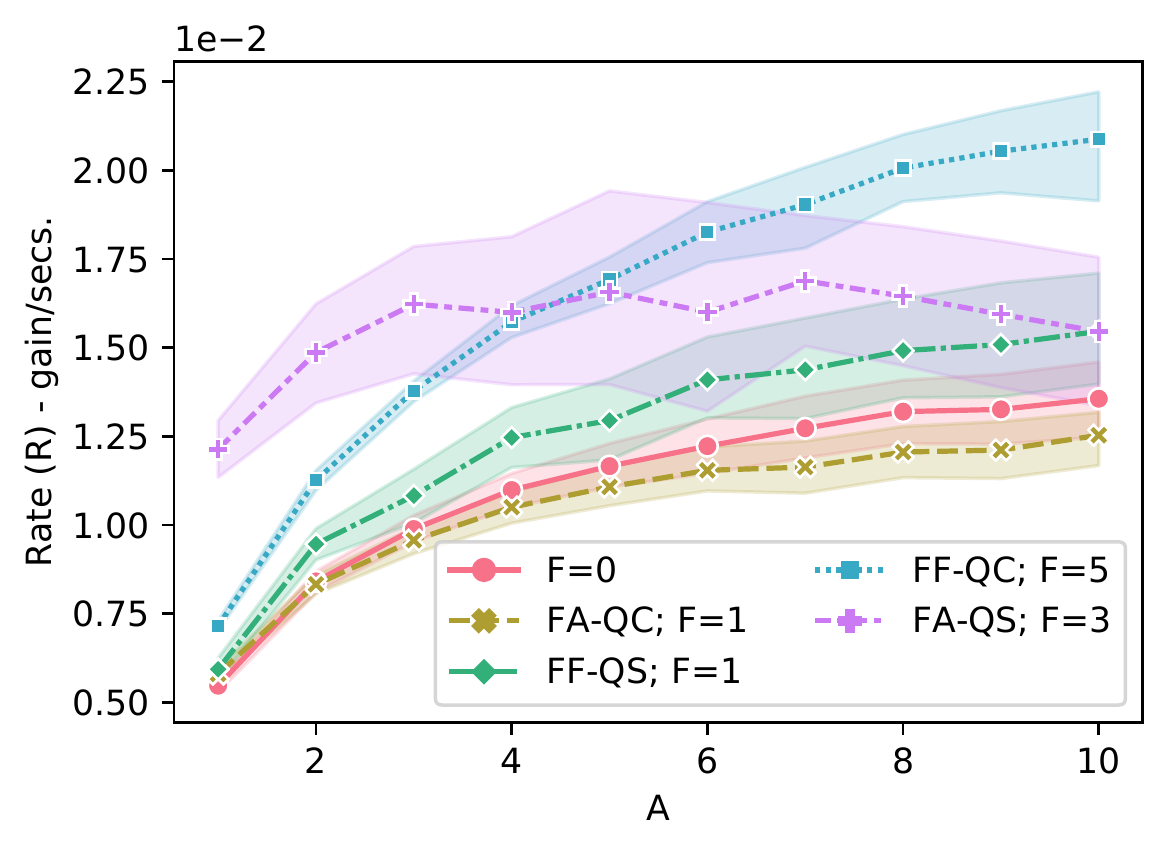}
    \caption{The Rate of Gain (R) by the number of Assessments (A) for the F value that yields the high rate of gain for each combination. FA-QC and FF-QS are clearly inferior to FF-QC and FA-QS, respectively. For clarity, only A$\le$10 is shown.}
    \label{fig:combo}
    \vspace{-5mm}
\end{figure}

To directly compare the different combinations of search strategy and mixed initiative, we have plotted the best performing combinations for each in Fig.~\ref{fig:combo}. Here we can see that the FA-QC (with $F=1$) combination is clearly inferior, while the FF-QS (with $F=1$) leads to a small increase over the baseline. More interestingly, we see that FA-QS (with $F=3$) outperforms the baselines and the other two combinations mentioned. However, FF-QC (with $F=5$) leads to the highest rate of gains overall, if the user is willing to assess five or more results per query. This suggests that there is no dominant strategy/approach but two competing combinations. Thus, for the remainder of our analysis, we focus on these two superior combinations: FF-QC ($F=5$), and FA-QS ($F=3$).

Table~\ref{tab:best_r} provides similar insights as described above, where we have listed the configurations that lead to the higher rate of gain for different strategies/approaches for three levels of gain. The table shows that as the amount of gain desired increases then the FF-QC combination results in the highest rate of gain.

\subsection{Query Length}
\begin{figure}[t]
    \vspace{-0.5cm}
    \subfloat[FF-QC; F=5]{\includegraphics[height=3.45cm]{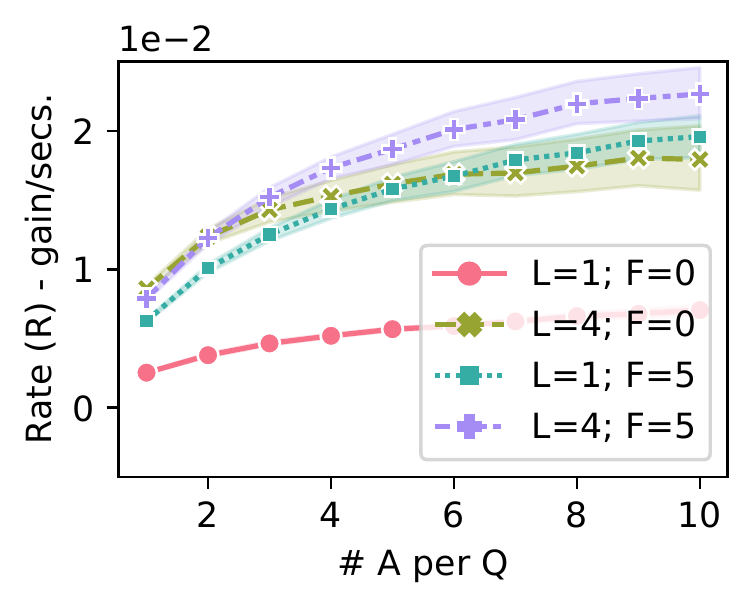}\label{fig:query-length-ff-qc}}
    \subfloat[FA-QS; F=3]{\includegraphics[height=3.45cm,trim=20 0 0 0, clip]{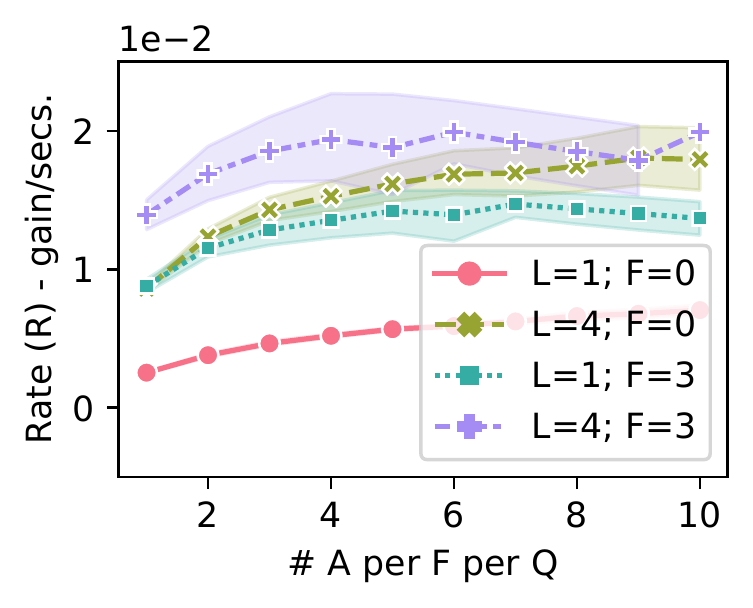}\label{fig:query-length-fa-qs}}
    \caption{The Rate of Gain (R) vs Assessments (A) for query lengths L=1 \& L=4 with no feedback (F=0) \& the F that yields the higher rate of gain. Left: Feedback First - Query Clarifications, Right: Feedback After - Query Suggestions. For clarity, only A$\le$10 is shown.
    }
    \label{fig:query-length}
    \vspace{-5mm}
\end{figure}

To explore our next research question on how the quality of queries influences the choice of strategy and mixed initiative approach taken, we examined how the rate of gain for each combination changed when we varied query length (and consequently the retrieval performance), see.
Fig.~\ref{fig:query-length}. In the plots, we can see that as query length increases from L=1 to L=4, the rate of gain also increases regardless of condition -- which is to be expected~\cite{belkin2003query_length}.

In Fig.~\ref{fig:query-length-ff-qc} we have plotted the rate of gain for FF-QC($F=5$). We can see that the increase in query length leads to a higher rate of gain. However, when compared to the no feedback condition, the rate of gain is similar when A is less than 3, but after providing feedback leads to higher rates of gain (when L=4).

A different story emerges in Fig.~\ref{fig:query-length-fa-qs} (right), where we have plotted the rate of gain for FA-QS ($F=3$). Here, when the length of the starting query is short (L=1), obtaining feedback from query suggestions leads to dramatic improvements in the rate of gain. However, if the starting query is longer (L=4), then the benefit of obtaining feedback via query suggestions leads to smaller increases in the rate of gain. Finally, as the number of assessments a user is willing to make increases, the benefit of feedback rounds from query suggestions diminishes and leads to a similar rate of gain as the no feedback baseline. Essentially, going deeper mitigates conversational interactions.

\vspace{-2mm}
\subsection{Cost of Conversational Turns}\label{sec:cost-changes} 
\begin{figure}[t]
    \vspace{-0.5cm}
    \subfloat[FF-QC; Changing $c_f$]{\includegraphics[height=3.45cm]{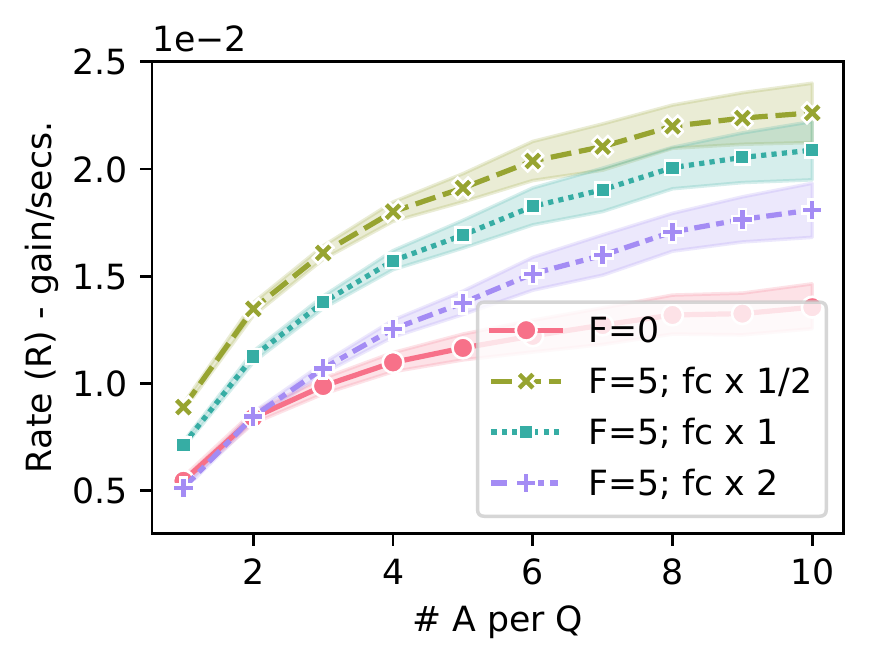}\label{fig:changing-costs-ff-qc-cf}}
    \subfloat[FF-QC; Changing $c_q$]{\includegraphics[height=3.45cm,trim=20 0 0 0, clip]{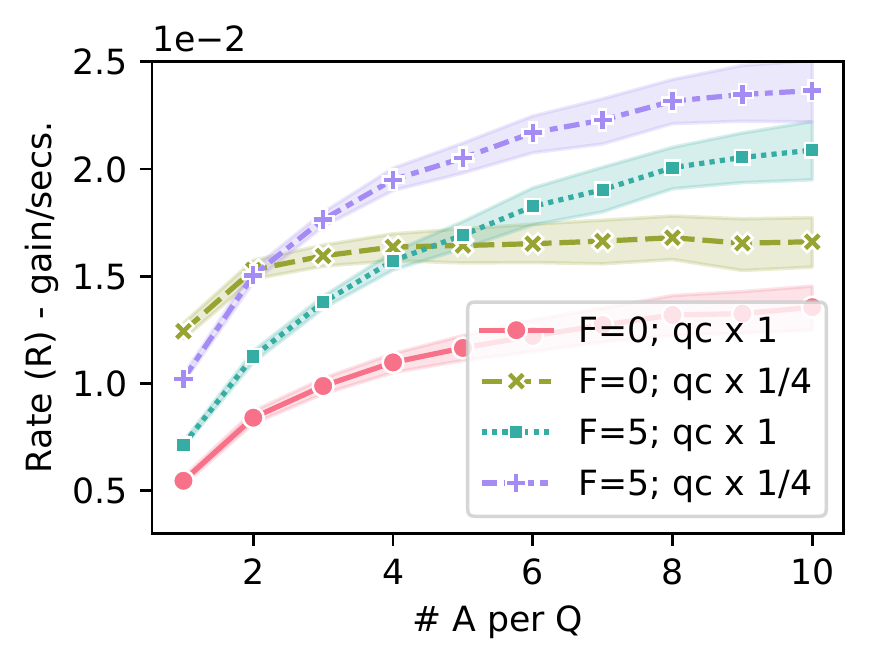}\label{fig:changing-costs-ff-qc-cq}}
    \vspace{-3mm}
    \subfloat[FA-QS; Changing $c_f$]{\includegraphics[height=3.45cm]{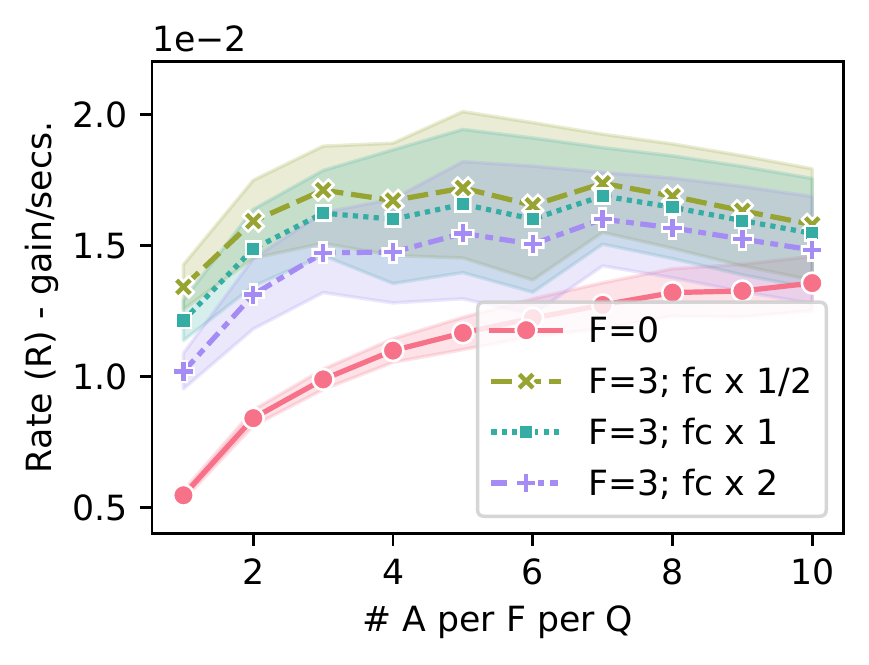}\label{fig:changing-costs-fa-qs-cf}}
    \subfloat[FA-QS; Changing $c_q$]{\includegraphics[height=3.45cm,trim=20 0 0 0, clip]{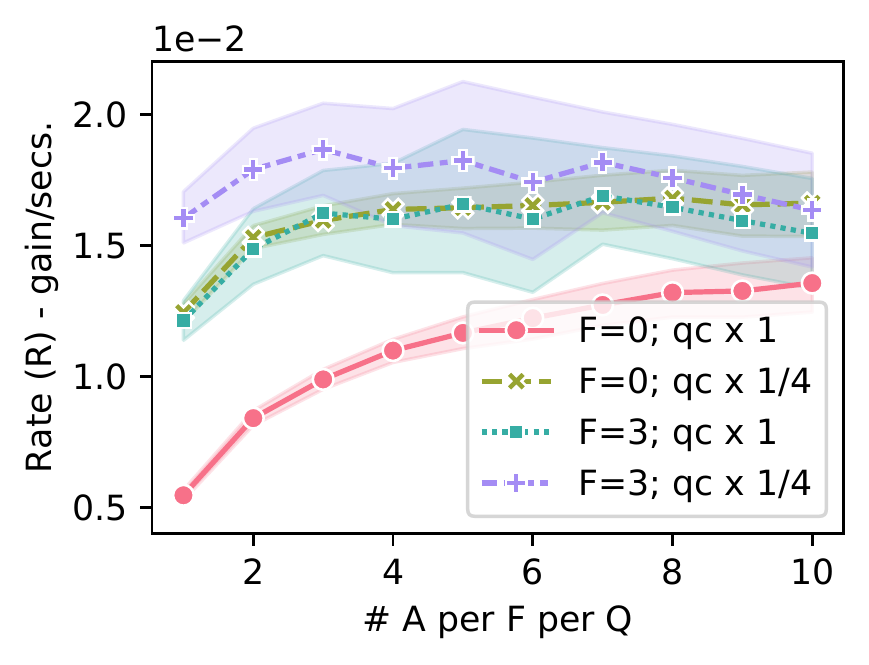}\label{fig:changing-costs-fa-qs-cq}}
    \caption{The Rate of Gain (R) vs Assessments (A) as the query cost of querying and the cost of feedback is varied. For clarity, only F$\le$5  and A$\le$10 is shown. 
    }
    \label{fig:changing-costs}
     \vspace{-5mm}
\end{figure}

To answer our final question, we explore whether changes to the costs affected the viability of the strategy or approach. So far we have used costs grounded by our user study, but what happens if the average cost associated with the different conversational turns changes? To explore how this might affect behaviours, we  varied the cost of feedback, in two ways: (i) by halving it and (ii) by doubling it. For FF-QC, Fig.~\ref{fig:changing-costs-ff-qc-cf} shows that as the feedback cost decreases it leads to a higher rate of gain. And, as the cost of feedback increases, we see that rate of gain decreases, making the combination less attractive when A is low. For FA-QS,  Fig.~\ref{fig:changing-costs-fa-qs-cf} shows that as feedback cost decreases, the rate of gain also increases, and this makes the combination worthwhile up until A is around 5-6 assessments. But when feedback cost increases, then the viability of the combination diminishes quickly. And, in fact, it eventually becomes worse than no feedback at all. 

In terms of changes to query cost, when we reduce the cost of querying, then the rate of gain for the baseline increases (as users reach relevant material sooner) -- and so we have updated the baselines in Fig.~\ref{fig:changing-costs-ff-qc-cq} and \ref{fig:changing-costs-fa-qs-cq}. For FF-QC, while previously providing clarifications resulted in a higher rate of gain, the decrease in query costs, means that FF-QC is only effective when A is less than 2, after that point re-querying results in a higher rate of gain. For FA-QS, the suggestions still result in a higher rate of gain than the no feedback baseline -- but the difference between the feedback and no feedback condition is considerably reduced -- and as A gets larger the differences between becomes smaller and smaller. Essentially, once queries become cheap enough, then issuing a series of queries, even if some are poor, is likely to lead to a higher rate of gain (as previously observed in~\cite{keskustalo2009test} during session search), rather than trying to refine the query through feedback. 

Regardless of the combination, we found that if assessment cost decreases then the rate of gain (R) increases, as less time is needed to extract relevant information, and conversely as the assessment cost increase then the rate of gain decreases. However, changing the cost of assessment didn't impact when to give feedback relative to the number of assessments (plots not shown).

\vspace{-1mm}
\section{Discussion and Future Work}
In this paper, we have explored how different CS strategies and different MI approaches combine in the context of a text based CSA where we have simulated CS sessions. In order to do so, we first built upon existing models of IIR to develop a model of the CS process which explicitly includes the core conversational concept of mixed initiative. From the model, we derived two different CS strategies, which have been previously observed in conversational settings.  While these strategies reduced the evaluation space, it is still largely intractable to explore all possible factors and so we focused on the most salient (i.e. number of A, F and Q, given the different conditions). 

With respect to the different conditions, we found that there was no dominant CS strategy and MI approach combination. However, we did observe that certain combinations were clearly inferior (e.g. FF-QS and FA-QC), while the choice of combination FA-QC led to higher rates of gain when A was lower, whereas for FF-QC higher rates of were observed when A was greater. Nonetheless, the viability of these combinations was dependent upon the initial query submitted, and the relative cost of giving feedback vs. the cost of querying. In sum, if (i) the length/quality of the initial queries increases, (ii) the cost of giving feedback increases, (iii) the cost of querying decreases, or (iv) a combination of, then providing feedback regardless of combination becomes less beneficial (resulting in a lower rate of gain), and it may even be detrimental where the rate of gain drops below the no feedback / non-conversational baseline. These findings begin to illuminate the complexities and trade-offs involved in conversational search, where it is clear that certain criteria need to be met for conversational search to be beneficial in terms of the rate of gain. 

It should be noted, however, that our findings need to be considered in context. We evaluated one particular type of CSA -- a chat/text based CSA like those proposed in~\cite{avula2017,Kaushik,Zamani2019} -- where we employed the traditional IR evaluation approach in a conversational setting.  We also used simulation based methodology so that we could begin to explore the large evaluation space (which would be near impossible to do so within a user study). Even so, we could only explore a subset of possibilities and focused on pure strategies with fixed rounds of feedback, etc.. Nonetheless, by evaluating and comparing pure strategies combined with the different mixed initiative approaches, we were still able to observe the strengths and weaknesses of the combinations and better understand the different trade-offs. 
In practice, however, it is clear that a mixture of different strategies and approaches will be employed and required to optimize the rate of the gain experienced during a CS session. As more interaction data becomes available from deployed CSAs it will be possible to instantiate more nuanced interaction models, and to evaluate other conversational search settings where the costs and gains vary. Clearly, this would change the pay-off dynamics associates with the different conversational turns -- and so evaluating different types of CSAs that, for example, try to surface relevant information directly would invariably lead to different strategies evolving. We have also made an assumption that CS should be as efficient as possible (following Grice's maxims of conversation~\cite{grice1975logic}) and that the users of CSAs and the CSAs will adapt/evolve to maximise the rate of gain (as per~Information Foraging Theory~\cite{DBLP:conf/chi/PirolliC95}). However, it is possible that the conversation itself has additional benefits leading to greater user satisfaction which may not be captured by focusing solely on gain, cost or rate measures. For example, in previous work, they found that asking a relevant clarification increased user satisfaction in voice-only conversations~\cite{DBLP:conf/sigir/KieselBSAH18} and so this may lead to other trade-offs emerging with satisfaction. Also, in this work, we solely relied on the TREC assessments. In a more realistic experimental setup, one could compute gain based on the amount of useful information given the agent's response. But, these are emerging challenges within the context of CS that need to be addressed through the development of more fine grained test collections before we can evaluate such scenarios.

In this paper, we have shown that the choice of search strategy and mixed initiative depends upon a number of factors: the quality of the starting query, the relative costs of querying vs. giving feedback, the number of results the user is willing to assess, and the amount of gain desired. While more work is needed to explore and investigate the effectiveness of different CSA configurations, the methods they used, and the strategies they employ, we have provided a model and framework for evaluating and simulating the conversational search process in an offline/batch setting. This will enable researchers to explore the complexities and trade-offs of design decisions before developing and deploying them in practice.

{
\partitle{Acknowledgements.}
    This work was supported in part by the NWO Innovational Research Incentives Scheme Vidi (016.Vidi.189.039), 
    the NWO Smart Culture - Big Data / Digital Humanities (314-99-301),
    the H2020-EU.3.4. - SOCIETAL CHALLENGES - Smart, Green And Integrated Transport (814961),
    and in part by the Center for Intelligent Information Retrieval.
    Any opinions, findings and conclusions or recommendations expressed in this material are those of the authors and do not necessarily reflect those of the sponsors. 
}

\newpage
\balance
\bibliographystyle{ACM-Reference-Format}
\bibliography{sigproc} 


\begin{thebibliography}{72}


\ifx \showCODEN    \undefined \def \showCODEN     #1{\unskip}     \fi
\ifx \showDOI      \undefined \def \showDOI       #1{#1}\fi
\ifx \showISBNx    \undefined \def \showISBNx     #1{\unskip}     \fi
\ifx \showISBNxiii \undefined \def \showISBNxiii  #1{\unskip}     \fi
\ifx \showISSN     \undefined \def \showISSN      #1{\unskip}     \fi
\ifx \showLCCN     \undefined \def \showLCCN      #1{\unskip}     \fi
\ifx \shownote     \undefined \def \shownote      #1{#1}          \fi
\ifx \showarticletitle \undefined \def \showarticletitle #1{#1}   \fi
\ifx \showURL      \undefined \def \showURL       {\relax}        \fi
\providecommand\bibfield[2]{#2}
\providecommand\bibinfo[2]{#2}
\providecommand\natexlab[1]{#1}
\providecommand\showeprint[2][]{arXiv:#2}

\bibitem[\protect\citeauthoryear{Aliannejadi, Kiseleva, Chuklin, Dalton, and
  Burtsev}{Aliannejadi et~al\mbox{.}}{2021}]%
        {aliannejadi21buidling}
\bibfield{author}{\bibinfo{person}{Mohammad Aliannejadi},
  \bibinfo{person}{Julia Kiseleva}, \bibinfo{person}{Aleksandr Chuklin},
  \bibinfo{person}{Jeffrey Dalton}, {and} \bibinfo{person}{Mikhail Burtsev}.}
  \bibinfo{year}{2021}\natexlab{}.
\newblock \showarticletitle{Building and Evaluating Open-Domain Dialogue
  Corpora \\ with Clarifying Questions}. In
  \bibinfo{booktitle}{\emph{{EMNLP}}}.
\newblock


\bibitem[\protect\citeauthoryear{Aliannejadi, Kiseleva, Chuklin, Dalton, and
  Burtsev}{Aliannejadi et~al\mbox{.}}{2020}]%
        {DBLP:journals/corr/abs-2009-11352}
\bibfield{author}{\bibinfo{person}{Mohammad Aliannejadi},
  \bibinfo{person}{Julia Kiseleva}, \bibinfo{person}{Aleksandr Chuklin},
  \bibinfo{person}{Jeff Dalton}, {and} \bibinfo{person}{Mikhail~S. Burtsev}.}
  \bibinfo{year}{2020}\natexlab{}.
\newblock \showarticletitle{ConvAI3: Generating Clarifying Questions for
  Open-Domain Dialogue Systems (ClariQ)}.
\newblock \bibinfo{journal}{\emph{CoRR}}  \bibinfo{volume}{abs/2009.11352}
  (\bibinfo{year}{2020}).
\newblock


\bibitem[\protect\citeauthoryear{Aliannejadi, Zamani, Crestani, and
  Croft}{Aliannejadi et~al\mbox{.}}{2019}]%
        {DBLP:conf/sigir/AliannejadiZCC19}
\bibfield{author}{\bibinfo{person}{Mohammad Aliannejadi},
  \bibinfo{person}{Hamed Zamani}, \bibinfo{person}{Fabio Crestani}, {and}
  \bibinfo{person}{W.~Bruce Croft}.} \bibinfo{year}{2019}\natexlab{}.
\newblock \showarticletitle{Asking Clarifying Questions in Open-Domain
  Information-Seeking Conversations}. In \bibinfo{booktitle}{\emph{{SIGIR}}}.
  \bibinfo{pages}{475--484}.
\newblock


\bibitem[\protect\citeauthoryear{Allen, Guinn, and Horvtz}{Allen
  et~al\mbox{.}}{1999}]%
        {allen1999mixed}
\bibfield{author}{\bibinfo{person}{JE Allen}, \bibinfo{person}{Curry~I Guinn},
  {and} \bibinfo{person}{Eric Horvtz}.} \bibinfo{year}{1999}\natexlab{}.
\newblock \showarticletitle{Mixed-Initiative Interaction}.
\newblock \bibinfo{journal}{\emph{{IEEE} Intell. Syst.}} \bibinfo{volume}{14},
  \bibinfo{number}{5} (\bibinfo{year}{1999}), \bibinfo{pages}{14--23}.
\newblock


\bibitem[\protect\citeauthoryear{Anand, Cavedon, Joho, Sanderson, and
  Stein}{Anand et~al\mbox{.}}{2019}]%
        {DBLP:journals/dagstuhl-reports/AnandCJSS19}
\bibfield{author}{\bibinfo{person}{Avishek Anand}, \bibinfo{person}{Lawrence
  Cavedon}, \bibinfo{person}{Hideo Joho}, \bibinfo{person}{Mark Sanderson},
  {and} \bibinfo{person}{Benno Stein}.} \bibinfo{year}{2019}\natexlab{}.
\newblock \showarticletitle{Conversational Search (Dagstuhl Seminar 19461)}.
\newblock \bibinfo{journal}{\emph{Dagstuhl Reports}} \bibinfo{volume}{9},
  \bibinfo{number}{11} (\bibinfo{year}{2019}), \bibinfo{pages}{34--83}.
\newblock


\bibitem[\protect\citeauthoryear{Avula}{Avula}{2017}]%
        {avula2017}
\bibfield{author}{\bibinfo{person}{Sandeep Avula}.}
  \bibinfo{year}{2017}\natexlab{}.
\newblock \showarticletitle{Searchbots: Using Chatbots in Collaborative
  Information-Seeking Tasks}. In \bibinfo{booktitle}{\emph{{SIGIR}}}.
  \bibinfo{pages}{1375}.
\newblock


\bibitem[\protect\citeauthoryear{Azzopardi}{Azzopardi}{2011}]%
        {DBLP:conf/sigir/Azzopardi11}
\bibfield{author}{\bibinfo{person}{Leif Azzopardi}.}
  \bibinfo{year}{2011}\natexlab{}.
\newblock \showarticletitle{The Economics in Interactive Information
  Retrieval}. In \bibinfo{booktitle}{\emph{{SIGIR}}}. \bibinfo{pages}{15--24}.
\newblock


\bibitem[\protect\citeauthoryear{Azzopardi, de~Rijke, and Balog}{Azzopardi
  et~al\mbox{.}}{2007}]%
        {DBLP:conf/sigir/AzzopardiRB07}
\bibfield{author}{\bibinfo{person}{Leif Azzopardi}, \bibinfo{person}{Maarten de
  Rijke}, {and} \bibinfo{person}{Krisztian Balog}.}
  \bibinfo{year}{2007}\natexlab{}.
\newblock \showarticletitle{Building Simulated Queries for Known-Item Topics:
  An Analysis using Six European Languages}. In
  \bibinfo{booktitle}{\emph{{SIGIR}}}. \bibinfo{pages}{455--462}.
\newblock


\bibitem[\protect\citeauthoryear{Azzopardi, Dubiel, Halvey, and
  Dalton}{Azzopardi et~al\mbox{.}}{2018a}]%
        {azzopardi2018conceptual}
\bibfield{author}{\bibinfo{person}{Leif Azzopardi}, \bibinfo{person}{Mateusz
  Dubiel}, \bibinfo{person}{Martin Halvey}, {and} \bibinfo{person}{Jeffery
  Dalton}.} \bibinfo{year}{2018}\natexlab{a}.
\newblock \showarticletitle{Conceptualizing Agent-Human Interactions during the
  Conversational Search Process}. In \bibinfo{booktitle}{\emph{CAIR}}.
\newblock


\bibitem[\protect\citeauthoryear{Azzopardi, Kelly, and Brennan}{Azzopardi
  et~al\mbox{.}}{2013}]%
        {DBLP:conf/sigir/AzzopardiKB13}
\bibfield{author}{\bibinfo{person}{Leif Azzopardi}, \bibinfo{person}{Diane
  Kelly}, {and} \bibinfo{person}{Kathy Brennan}.}
  \bibinfo{year}{2013}\natexlab{}.
\newblock \showarticletitle{How Query Cost Affects Search Behavior}. In
  \bibinfo{booktitle}{\emph{{SIGIR}}}. \bibinfo{pages}{23--32}.
\newblock


\bibitem[\protect\citeauthoryear{Azzopardi, Thomas, and Craswell}{Azzopardi
  et~al\mbox{.}}{2018b}]%
        {azzopardi2017ift}
\bibfield{author}{\bibinfo{person}{Leif Azzopardi}, \bibinfo{person}{Paul
  Thomas}, {and} \bibinfo{person}{Nick Craswell}.}
  \bibinfo{year}{2018}\natexlab{b}.
\newblock \showarticletitle{Measuring the Utility of Search Engine Result
  Pages: An Information Foraging Based Measure}. In
  \bibinfo{booktitle}{\emph{{SIGIR}}}. \bibinfo{pages}{605–614}.
\newblock


\bibitem[\protect\citeauthoryear{Azzopardi, Thomas, and Moffat}{Azzopardi
  et~al\mbox{.}}{2019}]%
        {DBLP:conf/sigir/AzzopardiTM19}
\bibfield{author}{\bibinfo{person}{Leif Azzopardi}, \bibinfo{person}{Paul
  Thomas}, {and} \bibinfo{person}{Alistair Moffat}.}
  \bibinfo{year}{2019}\natexlab{}.
\newblock \showarticletitle{cwl{\_}eval: An Evaluation Tool for Information
  Retrieval}. In \bibinfo{booktitle}{\emph{{SIGIR}}}.
  \bibinfo{pages}{1321--1324}.
\newblock


\bibitem[\protect\citeauthoryear{Baskaya, Keskustalo, and
  J{\"{a}}rvelin}{Baskaya et~al\mbox{.}}{2012}]%
        {Baskaya2012}
\bibfield{author}{\bibinfo{person}{Feza Baskaya}, \bibinfo{person}{Heikki
  Keskustalo}, {and} \bibinfo{person}{Kalervo J{\"{a}}rvelin}.}
  \bibinfo{year}{2012}\natexlab{}.
\newblock \showarticletitle{{Time Drives Interaction: Simulating Sessions in
  Diverse Searching Environments}}. In \bibinfo{booktitle}{\emph{{SIGIR}}}.
  \bibinfo{pages}{105--114}.
\newblock


\bibitem[\protect\citeauthoryear{Belkin, Cool, Stein, and Thiel}{Belkin
  et~al\mbox{.}}{1995}]%
        {belkin1995cases}
\bibfield{author}{\bibinfo{person}{Nicholas~J Belkin}, \bibinfo{person}{Colleen
  Cool}, \bibinfo{person}{Adelheit Stein}, {and} \bibinfo{person}{Ulrich
  Thiel}.} \bibinfo{year}{1995}\natexlab{}.
\newblock \showarticletitle{Cases, Scripts, and Information-Seeking Strategies:
  On the Design of Interactive Information Retrieval Systems}.
\newblock \bibinfo{journal}{\emph{Expert Syst. Appl.}} \bibinfo{volume}{9},
  \bibinfo{number}{3} (\bibinfo{year}{1995}), \bibinfo{pages}{379--395}.
\newblock


\bibitem[\protect\citeauthoryear{Belkin, Kelly, Kim, Kim, Lee, Muresan, Tang,
  Yuan, and Cool}{Belkin et~al\mbox{.}}{2003}]%
        {belkin2003query_length}
\bibfield{author}{\bibinfo{person}{Nicholas~J. Belkin}, \bibinfo{person}{D.
  Kelly}, \bibinfo{person}{G. Kim}, \bibinfo{person}{J.-Y. Kim},
  \bibinfo{person}{H.-J. Lee}, \bibinfo{person}{G. Muresan},
  \bibinfo{person}{M.-C. Tang}, \bibinfo{person}{X.-J. Yuan}, {and}
  \bibinfo{person}{C. Cool}.} \bibinfo{year}{2003}\natexlab{}.
\newblock \showarticletitle{Query Length in Interactive Information Retrieval}.
  In \bibinfo{booktitle}{\emph{{SIGIR}}}. \bibinfo{pages}{205–212}.
\newblock


\bibitem[\protect\citeauthoryear{Bosetti, Firmenich, Fern{\'{a}}ndez, Winckler,
  and Rossi}{Bosetti et~al\mbox{.}}{2017}]%
        {Bosetti2017}
\bibfield{author}{\bibinfo{person}{Gabriela Bosetti}, \bibinfo{person}{Sergio
  Firmenich}, \bibinfo{person}{Alejandro Fern{\'{a}}ndez},
  \bibinfo{person}{Marco Winckler}, {and} \bibinfo{person}{Gustavo Rossi}.}
  \bibinfo{year}{2017}\natexlab{}.
\newblock \showarticletitle{From Search Engines to Augmented Search Services:
  An End-User Development Approach}. In \bibinfo{booktitle}{\emph{{ICWE}}}.
  \bibinfo{pages}{115--133}.
\newblock


\bibitem[\protect\citeauthoryear{Braslavski, Savenkov, Agichtein, and
  Dubatovka}{Braslavski et~al\mbox{.}}{2017}]%
        {DBLP:conf/chiir/BraslavskiSAD17}
\bibfield{author}{\bibinfo{person}{Pavel Braslavski}, \bibinfo{person}{Denis
  Savenkov}, \bibinfo{person}{Eugene Agichtein}, {and} \bibinfo{person}{Alina
  Dubatovka}.} \bibinfo{year}{2017}\natexlab{}.
\newblock \showarticletitle{What Do You Mean Exactly?: Analyzing Clarification
  Questions in {CQA}}. In \bibinfo{booktitle}{\emph{{CHIIR}}}.
  \bibinfo{pages}{345--348}.
\newblock


\bibitem[\protect\citeauthoryear{Carterette}{Carterette}{2011}]%
        {Carterette2011}
\bibfield{author}{\bibinfo{person}{Ben Carterette}.}
  \bibinfo{year}{2011}\natexlab{}.
\newblock \showarticletitle{{System Effectiveness, User Models, and User
  Utility: A Conceptual Framework for Investigation}}. In
  \bibinfo{booktitle}{\emph{{SIGIR}}}. \bibinfo{pages}{903--912}.
\newblock


\bibitem[\protect\citeauthoryear{Croft}{Croft}{2019}]%
        {croft2019interaction}
\bibfield{author}{\bibinfo{person}{W.~Bruce Croft}.}
  \bibinfo{year}{2019}\natexlab{}.
\newblock \showarticletitle{The Importance of Interaction for Information
  Retrieval}. In \bibinfo{booktitle}{\emph{SIGIR}}. \bibinfo{pages}{1–2}.
\newblock


\bibitem[\protect\citeauthoryear{Croft and Thompson}{Croft and
  Thompson}{1987}]%
        {DBLP:journals/jasis/CroftT87}
\bibfield{author}{\bibinfo{person}{W.~Bruce Croft} {and} \bibinfo{person}{R.~H.
  Thompson}.} \bibinfo{year}{1987}\natexlab{}.
\newblock \showarticletitle{I\({}^{\mbox{3}}\)R: {A} New Approach to the Design
  of Document Retrieval Systems}.
\newblock \bibinfo{journal}{\emph{{J. Am. Soc. Inf. Sci.}}}
  \bibinfo{volume}{38}, \bibinfo{number}{6} (\bibinfo{year}{1987}),
  \bibinfo{pages}{389--404}.
\newblock


\bibitem[\protect\citeauthoryear{Culpepper, Diaz, and Smucker}{Culpepper
  et~al\mbox{.}}{2018}]%
        {DBLP:journals/sigir/Culpepper0S18}
\bibfield{author}{\bibinfo{person}{J.~Shane Culpepper},
  \bibinfo{person}{Fernando Diaz}, {and} \bibinfo{person}{Mark~D. Smucker}.}
  \bibinfo{year}{2018}\natexlab{}.
\newblock \showarticletitle{Research Frontiers in Information Retrieval: Report
  from the Third Strategic Workshop on Information Retrieval in Lorne {(SWIRL}
  2018)}.
\newblock \bibinfo{journal}{\emph{{SIGIR} Forum}} \bibinfo{volume}{52},
  \bibinfo{number}{1} (\bibinfo{year}{2018}), \bibinfo{pages}{34--90}.
\newblock


\bibitem[\protect\citeauthoryear{Dalton, Xiong, and Callan}{Dalton
  et~al\mbox{.}}{2019}]%
        {Dalton:2020:CAST}
\bibfield{author}{\bibinfo{person}{Jeffrey Dalton}, \bibinfo{person}{Chenyan
  Xiong}, {and} \bibinfo{person}{Jamie Callan}.}
  \bibinfo{year}{2019}\natexlab{}.
\newblock \showarticletitle{TREC CAsT 2019: The Conversational Assistance Track
  Overview}. In \bibinfo{booktitle}{\emph{{TREC}}}.
\newblock


\bibitem[\protect\citeauthoryear{Dubiel, Halvey, Azzopardi, Anderson, and
  Daronnat}{Dubiel et~al\mbox{.}}{2020}]%
        {Dubiel2020}
\bibfield{author}{\bibinfo{person}{M. Dubiel}, \bibinfo{person}{M. Halvey},
  \bibinfo{person}{L. Azzopardi}, \bibinfo{person}{D. Anderson}, {and}
  \bibinfo{person}{S. Daronnat}.} \bibinfo{year}{2020}\natexlab{}.
\newblock \showarticletitle{{Conversational Strategies: Impact on Search
  Performance in a Goal-Oriented Task Mateusz}}. In
  \bibinfo{booktitle}{\emph{CAIR}}.
\newblock


\bibitem[\protect\citeauthoryear{Grice}{Grice}{1975}]%
        {grice1975logic}
\bibfield{author}{\bibinfo{person}{Herbert~P Grice}.}
  \bibinfo{year}{1975}\natexlab{}.
\newblock \showarticletitle{Logic and conversation}.
\newblock In \bibinfo{booktitle}{\emph{Speech acts}}.
  \bibinfo{publisher}{Brill}, \bibinfo{pages}{41--58}.
\newblock


\bibitem[\protect\citeauthoryear{Hashemi, Zamani, and Croft}{Hashemi
  et~al\mbox{.}}{2020}]%
        {DBLP:conf/sigir/HashemiZC20}
\bibfield{author}{\bibinfo{person}{Helia Hashemi}, \bibinfo{person}{Hamed
  Zamani}, {and} \bibinfo{person}{W.~Bruce Croft}.}
  \bibinfo{year}{2020}\natexlab{}.
\newblock \showarticletitle{Guided Transformer: Leveraging Multiple External
  Sources for Representation Learning in Conversational Search}. In
  \bibinfo{booktitle}{\emph{{SIGIR}}}. \bibinfo{pages}{1131--1140}.
\newblock


\bibitem[\protect\citeauthoryear{He and Young}{He and Young}{2005}]%
        {DBLP:journals/csl/HeY05}
\bibfield{author}{\bibinfo{person}{Yulan He} {and} \bibinfo{person}{Steve~J.
  Young}.} \bibinfo{year}{2005}\natexlab{}.
\newblock \showarticletitle{Semantic Processing Using the Hidden Vector State
  Model}.
\newblock \bibinfo{journal}{\emph{Comput. Speech Lang.}} \bibinfo{volume}{19},
  \bibinfo{number}{1} (\bibinfo{year}{2005}), \bibinfo{pages}{85--106}.
\newblock


\bibitem[\protect\citeauthoryear{Huurnink, Hofmann, de~Rijke, and
  Bron}{Huurnink et~al\mbox{.}}{2010}]%
        {DBLP:conf/clef/HuurninkHRB10}
\bibfield{author}{\bibinfo{person}{Bouke Huurnink}, \bibinfo{person}{Katja
  Hofmann}, \bibinfo{person}{Maarten de Rijke}, {and} \bibinfo{person}{Marc
  Bron}.} \bibinfo{year}{2010}\natexlab{}.
\newblock \showarticletitle{Validating Query Simulators: An Experiment Using
  Commercial Searches and Purchases}. In \bibinfo{booktitle}{\emph{{CLEF}}}.
  \bibinfo{pages}{40--51}.
\newblock


\bibitem[\protect\citeauthoryear{J{\"a}rvelin and
  Kek{\"a}l{\"a}inen}{J{\"a}rvelin and Kek{\"a}l{\"a}inen}{2017}]%
        {jarvelin2017ir}
\bibfield{author}{\bibinfo{person}{Kalervo J{\"a}rvelin} {and}
  \bibinfo{person}{Jaana Kek{\"a}l{\"a}inen}.} \bibinfo{year}{2017}\natexlab{}.
\newblock \showarticletitle{IR Evaluation Methods for Retrieving Highly
  Relevant Documents}.
\newblock \bibinfo{journal}{\emph{SIGIR Forum}} \bibinfo{volume}{51},
  \bibinfo{number}{2} (\bibinfo{year}{2017}), \bibinfo{pages}{243--250}.
\newblock


\bibitem[\protect\citeauthoryear{J{\"a}rvelin, Price, Delcambre, and
  Nielsen}{J{\"a}rvelin et~al\mbox{.}}{2008}]%
        {kalervo2008sdcg}
\bibfield{author}{\bibinfo{person}{Kalervo J{\"a}rvelin},
  \bibinfo{person}{Susan~L. Price}, \bibinfo{person}{Lois M.~L. Delcambre},
  {and} \bibinfo{person}{Marianne~Lykke Nielsen}.}
  \bibinfo{year}{2008}\natexlab{}.
\newblock \showarticletitle{Discounted Cumulated Gain Based Evaluation of
  Multiple-Query IR Sessions}. In \bibinfo{booktitle}{\emph{ECIR}}.
  \bibinfo{pages}{4--15}.
\newblock


\bibitem[\protect\citeauthoryear{Johnston, Chen, Ehlen, Jung, Lieske, Reddy,
  Selfridge, Stoyanchev, Vasilieff, and Wilpon}{Johnston et~al\mbox{.}}{2014}]%
        {johnston-etal-2014-mva}
\bibfield{author}{\bibinfo{person}{Michael Johnston}, \bibinfo{person}{John
  Chen}, \bibinfo{person}{Patrick Ehlen}, \bibinfo{person}{Hyuckchul Jung},
  \bibinfo{person}{Jay Lieske}, \bibinfo{person}{Aarthi Reddy},
  \bibinfo{person}{Ethan Selfridge}, \bibinfo{person}{Svetlana Stoyanchev},
  \bibinfo{person}{Brant Vasilieff}, {and} \bibinfo{person}{Jay Wilpon}.}
  \bibinfo{year}{2014}\natexlab{}.
\newblock \showarticletitle{{MVA}: The Multimodal Virtual Assistant}. In
  \bibinfo{booktitle}{\emph{{{SIGDIAL}}}}. \bibinfo{pages}{257--259}.
\newblock


\bibitem[\protect\citeauthoryear{Jordan, Watters, and Gao}{Jordan
  et~al\mbox{.}}{2006}]%
        {DBLP:conf/jcdl/JordanWG06}
\bibfield{author}{\bibinfo{person}{Chris Jordan}, \bibinfo{person}{Carolyn~R.
  Watters}, {and} \bibinfo{person}{Qigang Gao}.}
  \bibinfo{year}{2006}\natexlab{}.
\newblock \showarticletitle{Using Controlled Query Generation to Evaluate Blind
  Relevance Feedback Algorithms}. In \bibinfo{booktitle}{\emph{{JCDL}}}.
  \bibinfo{pages}{286--295}.
\newblock


\bibitem[\protect\citeauthoryear{Kaushik, Ramachandra, and Jones}{Kaushik
  et~al\mbox{.}}{[n. d.]}]%
        {Kaushik}
\bibfield{author}{\bibinfo{person}{Abhishek Kaushik},
  \bibinfo{person}{Vishal~Bhat Ramachandra}, {and} \bibinfo{person}{Gareth~J.F.
  Jones}.} \bibinfo{year}{[n. d.]}\natexlab{}.
\newblock \showarticletitle{{An Interface for Agent Supported Conversational
  Search}}. In \bibinfo{booktitle}{\emph{{CHIIR}}}. \bibinfo{pages}{452--456}.
\newblock


\bibitem[\protect\citeauthoryear{Keskustalo, J{\"{a}}rvelin, and
  Pirkola}{Keskustalo et~al\mbox{.}}{2008}]%
        {DBLP:journals/ir/KeskustaloJP08}
\bibfield{author}{\bibinfo{person}{Heikki Keskustalo}, \bibinfo{person}{Kalervo
  J{\"{a}}rvelin}, {and} \bibinfo{person}{Ari Pirkola}.}
  \bibinfo{year}{2008}\natexlab{}.
\newblock \showarticletitle{Evaluating the Effectiveness of Relevance Feedback
  based on a User Simulation Model: Effects of a user Scenario on Cumulated
  Gain Value}.
\newblock \bibinfo{journal}{\emph{Inf. Retr.}} \bibinfo{volume}{11},
  \bibinfo{number}{3} (\bibinfo{year}{2008}), \bibinfo{pages}{209--228}.
\newblock


\bibitem[\protect\citeauthoryear{Keskustalo, J{\"a}rvelin, Pirkola, Sharma, and
  Lykke}{Keskustalo et~al\mbox{.}}{2009}]%
        {keskustalo2009test}
\bibfield{author}{\bibinfo{person}{Heikki Keskustalo}, \bibinfo{person}{Kalervo
  J{\"a}rvelin}, \bibinfo{person}{Ari Pirkola}, \bibinfo{person}{Tarun Sharma},
  {and} \bibinfo{person}{Marianne Lykke}.} \bibinfo{year}{2009}\natexlab{}.
\newblock \showarticletitle{Test Collection-based IR Evaluation Needs Extension
  Toward Sessions--A Case of Extremely Short Queries}. In
  \bibinfo{booktitle}{\emph{{AIRS}}}. \bibinfo{pages}{63--74}.
\newblock


\bibitem[\protect\citeauthoryear{Kiesel, Bahrami, Stein, Anand, and
  Hagen}{Kiesel et~al\mbox{.}}{2018}]%
        {DBLP:conf/sigir/KieselBSAH18}
\bibfield{author}{\bibinfo{person}{Johannes Kiesel}, \bibinfo{person}{Arefeh
  Bahrami}, \bibinfo{person}{Benno Stein}, \bibinfo{person}{Avishek Anand},
  {and} \bibinfo{person}{Matthias Hagen}.} \bibinfo{year}{2018}\natexlab{}.
\newblock \showarticletitle{Toward Voice Query Clarification}. In
  \bibinfo{booktitle}{\emph{{SIGIR}}}. \bibinfo{pages}{1257--1260}.
\newblock


\bibitem[\protect\citeauthoryear{Kiseleva, Williams, Hassan~Awadallah, Crook,
  Zitouni, and Anastasakos}{Kiseleva et~al\mbox{.}}{2016}]%
        {kiseleva2016predicting}
\bibfield{author}{\bibinfo{person}{Julia Kiseleva}, \bibinfo{person}{Kyle
  Williams}, \bibinfo{person}{Ahmed Hassan~Awadallah}, \bibinfo{person}{Aidan~C
  Crook}, \bibinfo{person}{Imed Zitouni}, {and} \bibinfo{person}{Tasos
  Anastasakos}.} \bibinfo{year}{2016}\natexlab{}.
\newblock \showarticletitle{Predicting User Satisfaction with Intelligent
  Assistants}. In \bibinfo{booktitle}{\emph{SIGIR}}. \bibinfo{pages}{45--54}.
\newblock


\bibitem[\protect\citeauthoryear{Krasakis, Aliannejadi, Voskarides, and
  Kanoulas}{Krasakis et~al\mbox{.}}{2020}]%
        {DBLP:conf/ictir/KrasakisAVK20}
\bibfield{author}{\bibinfo{person}{Antonios~Minas Krasakis},
  \bibinfo{person}{Mohammad Aliannejadi}, \bibinfo{person}{Nikos Voskarides},
  {and} \bibinfo{person}{Evangelos Kanoulas}.} \bibinfo{year}{2020}\natexlab{}.
\newblock \showarticletitle{Analysing the Effect of Clarifying Questions on
  Document Ranking in Conversational Search}. In
  \bibinfo{booktitle}{\emph{{ICTIR}}}. \bibinfo{pages}{129--132}.
\newblock


\bibitem[\protect\citeauthoryear{Lipani, Carterette, and Yilmaz}{Lipani
  et~al\mbox{.}}{2021}]%
        {lipani2021evaluating_css}
\bibfield{author}{\bibinfo{person}{Aldo Lipani}, \bibinfo{person}{Ben
  Carterette}, {and} \bibinfo{person}{Emine Yilmaz}.}
  \bibinfo{year}{2021}\natexlab{}.
\newblock \showarticletitle{How Am I Doing?: Evaluating Conversational Search
  Systems Offline}.
\newblock \bibinfo{journal}{\emph{ACM Trans. Inf. Syst.}} \bibinfo{volume}{39},
  \bibinfo{number}{4} (\bibinfo{year}{2021}).
\newblock


\bibitem[\protect\citeauthoryear{Lotze, Klut, Aliannejadi, and Kanoulas}{Lotze
  et~al\mbox{.}}{2021}]%
        {DBLP:journals/corr/abs-2103-06192}
\bibfield{author}{\bibinfo{person}{Tom Lotze}, \bibinfo{person}{Stefan Klut},
  \bibinfo{person}{Mohammad Aliannejadi}, {and} \bibinfo{person}{Evangelos
  Kanoulas}.} \bibinfo{year}{2021}\natexlab{}.
\newblock \showarticletitle{Ranking Clarifying Questions Based on Predicted
  User Engagement}.
\newblock \bibinfo{journal}{\emph{CoRR}}  \bibinfo{volume}{abs/2103.06192}
  (\bibinfo{year}{2021}).
\newblock


\bibitem[\protect\citeauthoryear{Marchionini}{Marchionini}{1997}]%
        {Marchionini1997}
\bibfield{author}{\bibinfo{person}{Gary Marchionini}.}
  \bibinfo{year}{1997}\natexlab{}.
\newblock \bibinfo{booktitle}{\emph{{Information Seeking in Electronic
  Environments}}}.
\newblock \bibinfo{publisher}{Cambridge University Press},
  \bibinfo{address}{USA}.
\newblock
\showISBNx{0521586747}


\bibitem[\protect\citeauthoryear{Maxwell and Azzopardi}{Maxwell and
  Azzopardi}{2016}]%
        {maxwell2016agents}
\bibfield{author}{\bibinfo{person}{David Maxwell} {and} \bibinfo{person}{Leif
  Azzopardi}.} \bibinfo{year}{2016}\natexlab{}.
\newblock \showarticletitle{Agents, simulated users and humans: An analysis of
  performance and behaviour}. In \bibinfo{booktitle}{\emph{{CIKM}}}.
  \bibinfo{pages}{731--740}.
\newblock


\bibitem[\protect\citeauthoryear{Maxwell, Azzopardi, J{\"{a}}rvelin, and
  Keskustalo}{Maxwell et~al\mbox{.}}{2015}]%
        {Maxwell2015}
\bibfield{author}{\bibinfo{person}{David Maxwell}, \bibinfo{person}{Leif
  Azzopardi}, \bibinfo{person}{Kalervo J{\"{a}}rvelin}, {and}
  \bibinfo{person}{Heikki Keskustalo}.} \bibinfo{year}{2015}\natexlab{}.
\newblock \showarticletitle{{Searching and Stopping: An Analysis of Stopping
  Rules and Strategies}}. In \bibinfo{booktitle}{\emph{{CIKM}}}.
  \bibinfo{pages}{313--322}.
\newblock


\bibitem[\protect\citeauthoryear{McTear}{McTear}{2002}]%
        {McTear2002}
\bibfield{author}{\bibinfo{person}{Michael~F. McTear}.}
  \bibinfo{year}{2002}\natexlab{}.
\newblock \showarticletitle{Spoken Dialogue Technology: Enabling the
  Conversational User Interface}.
\newblock \bibinfo{journal}{\emph{ACM Comput. Surv.}} \bibinfo{volume}{34},
  \bibinfo{number}{1} (\bibinfo{year}{2002}), \bibinfo{pages}{90–169}.
\newblock


\bibitem[\protect\citeauthoryear{Moffat, Thomas, and Scholer}{Moffat
  et~al\mbox{.}}{2013}]%
        {DBLP:conf/cikm/MoffatTS13}
\bibfield{author}{\bibinfo{person}{Alistair Moffat}, \bibinfo{person}{Paul
  Thomas}, {and} \bibinfo{person}{Falk Scholer}.}
  \bibinfo{year}{2013}\natexlab{}.
\newblock \showarticletitle{Users versus Models: What Observation Tells us
  about Effectiveness Metrics}. In \bibinfo{booktitle}{\emph{{CIKM}}}.
  \bibinfo{pages}{659--668}.
\newblock


\bibitem[\protect\citeauthoryear{Moffat and Zobel}{Moffat and Zobel}{2008}]%
        {Moffat2008}
\bibfield{author}{\bibinfo{person}{Alistair Moffat} {and}
  \bibinfo{person}{Justin Zobel}.} \bibinfo{year}{2008}\natexlab{}.
\newblock \showarticletitle{{Rank-Biased Precision for Measurement of Retrieval
  Effectiveness}}.
\newblock \bibinfo{journal}{\emph{ACM Trans. on Inf. Sys.}}
  \bibinfo{volume}{27}, \bibinfo{number}{1} (\bibinfo{year}{2008}),
  \bibinfo{pages}{2:1--2:27}.
\newblock


\bibitem[\protect\citeauthoryear{Oddy}{Oddy}{1977}]%
        {Oddy1977}
\bibfield{author}{\bibinfo{person}{R.N. Oddy}.}
  \bibinfo{year}{1977}\natexlab{}.
\newblock \showarticletitle{Information Retrieval Through Man‐Machine
  Dialogue}.
\newblock \bibinfo{journal}{\emph{J. Documentation}} \bibinfo{volume}{33},
  \bibinfo{number}{1} (\bibinfo{year}{1977}), \bibinfo{pages}{1--14}.
\newblock
\showISSN{0022-0418}


\bibitem[\protect\citeauthoryear{{Pieraccini}, {Tzoukermann}, {Gorelov},
  {Gauvain}, {Levin}, {Lee}, and {Wilpon}}{{Pieraccini} et~al\mbox{.}}{1992}]%
        {225939}
\bibfield{author}{\bibinfo{person}{Roberto {Pieraccini}},
  \bibinfo{person}{Evelyne {Tzoukermann}}, \bibinfo{person}{Z. {Gorelov}},
  \bibinfo{person}{Jean-Luc {Gauvain}}, \bibinfo{person}{Esther {Levin}},
  \bibinfo{person}{Chin-Hui {Lee}}, {and} \bibinfo{person}{Jay {Wilpon}}.}
  \bibinfo{year}{1992}\natexlab{}.
\newblock \showarticletitle{A Speech Understanding System based on Statistical
  Representation of Semantics}. In \bibinfo{booktitle}{\emph{ICASSP}}.
  \bibinfo{pages}{193--196}.
\newblock


\bibitem[\protect\citeauthoryear{Pirolli and Card}{Pirolli and Card}{1995}]%
        {DBLP:conf/chi/PirolliC95}
\bibfield{author}{\bibinfo{person}{Peter Pirolli} {and}
  \bibinfo{person}{Stuart~K. Card}.} \bibinfo{year}{1995}\natexlab{}.
\newblock \showarticletitle{Information Foraging in Information Access
  Environments}. In \bibinfo{booktitle}{\emph{{CHI}}}. \bibinfo{pages}{51--58}.
\newblock


\bibitem[\protect\citeauthoryear{Radlinski and Craswell}{Radlinski and
  Craswell}{2017}]%
        {DBLP:conf/chiir/RadlinskiC17}
\bibfield{author}{\bibinfo{person}{Filip Radlinski} {and} \bibinfo{person}{Nick
  Craswell}.} \bibinfo{year}{2017}\natexlab{}.
\newblock \showarticletitle{A Theoretical Framework for Conversational Search}.
  In \bibinfo{booktitle}{\emph{{CHIIR}}}. \bibinfo{pages}{117--126}.
\newblock


\bibitem[\protect\citeauthoryear{Rao and Daum{\'{e}}}{Rao and
  Daum{\'{e}}}{2018}]%
        {DBLP:conf/acl/DaumeR18}
\bibfield{author}{\bibinfo{person}{Sudha Rao} {and} \bibinfo{person}{Hal
  Daum{\'{e}}}.} \bibinfo{year}{2018}\natexlab{}.
\newblock \showarticletitle{Learning to Ask Good Questions: Ranking
  Clarification Questions using Neural Expected Value of Perfect Information}.
  In \bibinfo{booktitle}{\emph{{ACL} {(1)}}}. \bibinfo{pages}{2736--2745}.
\newblock


\bibitem[\protect\citeauthoryear{Sahib, Tombros, and Stockman}{Sahib
  et~al\mbox{.}}{2012}]%
        {Sahib2012}
\bibfield{author}{\bibinfo{person}{Nuzhah~Gooda Sahib},
  \bibinfo{person}{Anastasios Tombros}, {and} \bibinfo{person}{Tony Stockman}.}
  \bibinfo{year}{2012}\natexlab{}.
\newblock \showarticletitle{{A Comparative Analysis of the Information-Seeking
  Behavior of Visually Impaired and Sighted Searchers Nuzhah}}.
\newblock \bibinfo{journal}{\emph{J. Assoc. Inf. Sci. Technol.}}
  \bibinfo{volume}{63}, \bibinfo{number}{2} (\bibinfo{year}{2012}),
  \bibinfo{pages}{377--391}.
\newblock


\bibitem[\protect\citeauthoryear{Sekulic, Aliannejadi, and Crestani}{Sekulic
  et~al\mbox{.}}{2021}]%
        {DBLP:conf/ecir/SekulicAC21}
\bibfield{author}{\bibinfo{person}{Ivan Sekulic}, \bibinfo{person}{Mohammad
  Aliannejadi}, {and} \bibinfo{person}{Fabio Crestani}.}
  \bibinfo{year}{2021}\natexlab{}.
\newblock \showarticletitle{User Engagement Prediction for Clarification in
  Search}. In \bibinfo{booktitle}{\emph{{ECIR}}}. \bibinfo{pages}{619--633}.
\newblock


\bibitem[\protect\citeauthoryear{Sitter and Stein}{Sitter and Stein}{1992}]%
        {SITTER1992165}
\bibfield{author}{\bibinfo{person}{Stefan Sitter} {and}
  \bibinfo{person}{Adelheit Stein}.} \bibinfo{year}{1992}\natexlab{}.
\newblock \showarticletitle{Modeling the Illocutionary Aspects of
  Information-Seeking Dialogues}.
\newblock \bibinfo{journal}{\emph{Inf. Process. Manag.}} \bibinfo{volume}{28},
  \bibinfo{number}{2} (\bibinfo{year}{1992}), \bibinfo{pages}{165--180}.
\newblock
\showISSN{0306-4573}


\bibitem[\protect\citeauthoryear{Smucker and Clarke}{Smucker and
  Clarke}{2012}]%
        {DBLP:conf/sigir/SmuckerC12}
\bibfield{author}{\bibinfo{person}{Mark~D. Smucker} {and}
  \bibinfo{person}{Charles L.~A. Clarke}.} \bibinfo{year}{2012}\natexlab{}.
\newblock \showarticletitle{Time-based Calibration of Effectiveness Measures}.
  In \bibinfo{booktitle}{\emph{{SIGIR}}}. \bibinfo{pages}{95--104}.
\newblock


\bibitem[\protect\citeauthoryear{Spina, Trippas, Cavedon, and Sanderson}{Spina
  et~al\mbox{.}}{2017}]%
        {Spina:2017}
\bibfield{author}{\bibinfo{person}{Damiano Spina}, \bibinfo{person}{Johanne~R.
  Trippas}, \bibinfo{person}{Lawrence Cavedon}, {and} \bibinfo{person}{Mark
  Sanderson}.} \bibinfo{year}{2017}\natexlab{}.
\newblock \showarticletitle{Extracting Audio Summaries to Support Effective
  Spoken Document Search}.
\newblock \bibinfo{journal}{\emph{J. Assoc. Inf. Sci. Technol.}}
  \bibinfo{volume}{68}, \bibinfo{number}{9} (\bibinfo{year}{2017}),
  \bibinfo{pages}{2101–2115}.
\newblock


\bibitem[\protect\citeauthoryear{Thomas, Moffat, Bailey, and Scholer}{Thomas
  et~al\mbox{.}}{2014}]%
        {thomas2014modeling}
\bibfield{author}{\bibinfo{person}{Paul Thomas}, \bibinfo{person}{Alistair
  Moffat}, \bibinfo{person}{Peter Bailey}, {and} \bibinfo{person}{Falk
  Scholer}.} \bibinfo{year}{2014}\natexlab{}.
\newblock \showarticletitle{Modeling Decision Points in User Search Behavior}.
  In \bibinfo{booktitle}{\emph{{IIiX}}}. \bibinfo{pages}{239--242}.
\newblock


\bibitem[\protect\citeauthoryear{Trippas, Spina, Cavedon, Joho, and
  Sanderson}{Trippas et~al\mbox{.}}{2018a}]%
        {Trippas2018}
\bibfield{author}{\bibinfo{person}{Johanne~R. Trippas},
  \bibinfo{person}{Damiano Spina}, \bibinfo{person}{Lawrence Cavedon},
  \bibinfo{person}{Hideo Joho}, {and} \bibinfo{person}{Mark Sanderson}.}
  \bibinfo{year}{2018}\natexlab{a}.
\newblock \showarticletitle{{Informing the Design of Spoken Conversational
  Search}}. In \bibinfo{booktitle}{\emph{{CHIIR}}}. \bibinfo{pages}{32--41}.
\newblock
\showISBNx{9781450349253}


\bibitem[\protect\citeauthoryear{Trippas, Spina, Cavedon, Joho, and
  Sanderson}{Trippas et~al\mbox{.}}{2018b}]%
        {DBLP:conf/chiir/TrippasSCJS18}
\bibfield{author}{\bibinfo{person}{Johanne~R. Trippas},
  \bibinfo{person}{Damiano Spina}, \bibinfo{person}{Lawrence Cavedon},
  \bibinfo{person}{Hideo Joho}, {and} \bibinfo{person}{Mark Sanderson}.}
  \bibinfo{year}{2018}\natexlab{b}.
\newblock \showarticletitle{Informing the Design of Spoken Conversational
  Search: Perspective Paper}. In \bibinfo{booktitle}{\emph{{CHIIR}}}.
  \bibinfo{pages}{32--41}.
\newblock


\bibitem[\protect\citeauthoryear{Trippas, Spina, Thomas, Sanderson, Joho, and
  Cavedon}{Trippas et~al\mbox{.}}{2020}]%
        {Trippas2020}
\bibfield{author}{\bibinfo{person}{Johanne~R. Trippas},
  \bibinfo{person}{Damiano Spina}, \bibinfo{person}{Paul Thomas},
  \bibinfo{person}{Mark Sanderson}, \bibinfo{person}{Hideo Joho}, {and}
  \bibinfo{person}{Lawrence Cavedon}.} \bibinfo{year}{2020}\natexlab{}.
\newblock \showarticletitle{{Towards a Model for Spoken Conversational
  Search}}.
\newblock \bibinfo{journal}{\emph{Inf. Process. Manag.}} \bibinfo{volume}{57},
  \bibinfo{number}{2} (\bibinfo{year}{2020}), \bibinfo{pages}{102162}.
\newblock
\showISSN{03064573}


\bibitem[\protect\citeauthoryear{Vakulenko, Revoredo, Di~Ciccio, and
  de~Rijke}{Vakulenko et~al\mbox{.}}{2019}]%
        {vakulenko2019qrfa}
\bibfield{author}{\bibinfo{person}{Svitlana Vakulenko}, \bibinfo{person}{Kate
  Revoredo}, \bibinfo{person}{Claudio Di~Ciccio}, {and}
  \bibinfo{person}{Maarten de Rijke}.} \bibinfo{year}{2019}\natexlab{}.
\newblock \showarticletitle{QRFA: A Data-Driven Model of Information Seeking
  Dialogues}. In \bibinfo{booktitle}{\emph{{ECIR}}}. \bibinfo{pages}{541--557}.
\newblock


\bibitem[\protect\citeauthoryear{Vtyurina, Savenkov, Agichtein, and
  Clarke}{Vtyurina et~al\mbox{.}}{2017}]%
        {DBLP:conf/chi/VtyurinaSAC17}
\bibfield{author}{\bibinfo{person}{Alexandra Vtyurina}, \bibinfo{person}{Denis
  Savenkov}, \bibinfo{person}{Eugene Agichtein}, {and} \bibinfo{person}{Charles
  L.~A. Clarke}.} \bibinfo{year}{2017}\natexlab{}.
\newblock \showarticletitle{Exploring Conversational Search With Humans,
  Assistants, and Wizards}. In \bibinfo{booktitle}{\emph{{CHI} Extended
  Abstracts}}. \bibinfo{pages}{2187--2193}.
\newblock


\bibitem[\protect\citeauthoryear{Wadhwa and Zamani}{Wadhwa and Zamani}{2021}]%
        {Wadhwa2021}
\bibfield{author}{\bibinfo{person}{Somin Wadhwa} {and} \bibinfo{person}{Hamed
  Zamani}.} \bibinfo{year}{2021}\natexlab{}.
\newblock \showarticletitle{Towards System-Initiative Conversational
  Information Seeking}. In \bibinfo{booktitle}{\emph{{DESIRES}}}.
\newblock


\bibitem[\protect\citeauthoryear{Walker, Passonneau, and Boland}{Walker
  et~al\mbox{.}}{2001}]%
        {DBLP:conf/acl/WalkerPB01}
\bibfield{author}{\bibinfo{person}{Marilyn~A. Walker},
  \bibinfo{person}{Rebecca~J. Passonneau}, {and} \bibinfo{person}{Julie~E.
  Boland}.} \bibinfo{year}{2001}\natexlab{}.
\newblock \showarticletitle{Quantitative and Qualitative Evaluation of Darpa
  Communicator Spoken Dialogue Systems}. In \bibinfo{booktitle}{\emph{{ACL}}}.
  \bibinfo{pages}{515--522}.
\newblock


\bibitem[\protect\citeauthoryear{Wambua, Raimondo, Boger, Polgar, Chinaei, and
  Rudzicz}{Wambua et~al\mbox{.}}{2018}]%
        {Wambua2018}
\bibfield{author}{\bibinfo{person}{Muuo Wambua}, \bibinfo{person}{Stefania
  Raimondo}, \bibinfo{person}{Jennifer Boger}, \bibinfo{person}{Jan Polgar},
  \bibinfo{person}{Hamidreza Chinaei}, {and} \bibinfo{person}{Frank Rudzicz}.}
  \bibinfo{year}{2018}\natexlab{}.
\newblock \showarticletitle{{Interactive Search through Iterative Refinement}}.
  In \bibinfo{booktitle}{\emph{{CAIR}}}.
\newblock


\bibitem[\protect\citeauthoryear{Yan, Song, and Wu}{Yan et~al\mbox{.}}{2016}]%
        {DBLP:conf/sigir/YanSW16}
\bibfield{author}{\bibinfo{person}{Rui Yan}, \bibinfo{person}{Yiping Song},
  {and} \bibinfo{person}{Hua Wu}.} \bibinfo{year}{2016}\natexlab{}.
\newblock \showarticletitle{Learning to Respond with Deep Neural Networks for
  Retrieval-Based Human-Computer Conversation System}. In
  \bibinfo{booktitle}{\emph{{SIGIR}}}. \bibinfo{pages}{55--64}.
\newblock


\bibitem[\protect\citeauthoryear{Yang, Qiu, Qu, Guo, Zhang, Croft, Huang, and
  Chen}{Yang et~al\mbox{.}}{2018}]%
        {DBLP:conf/sigir/YangQQGZCHC18}
\bibfield{author}{\bibinfo{person}{Liu Yang}, \bibinfo{person}{Minghui Qiu},
  \bibinfo{person}{Chen Qu}, \bibinfo{person}{Jiafeng Guo},
  \bibinfo{person}{Yongfeng Zhang}, \bibinfo{person}{W.~Bruce Croft},
  \bibinfo{person}{Jun Huang}, {and} \bibinfo{person}{Haiqing Chen}.}
  \bibinfo{year}{2018}\natexlab{}.
\newblock \showarticletitle{Response Ranking with Deep Matching Networks and
  External Knowledge in Information-seeking Conversation Systems}. In
  \bibinfo{booktitle}{\emph{{SIGIR}}}. \bibinfo{pages}{245--254}.
\newblock


\bibitem[\protect\citeauthoryear{Zamani and Craswell}{Zamani and
  Craswell}{2020}]%
        {Zamani2019}
\bibfield{author}{\bibinfo{person}{Hamed Zamani} {and} \bibinfo{person}{Nick
  Craswell}.} \bibinfo{year}{2020}\natexlab{}.
\newblock \showarticletitle{Macaw: An Extensible Conversational Information
  Seeking Platform}. In \bibinfo{booktitle}{\emph{{SIGIR}}}.
  \bibinfo{pages}{2193–2196}.
\newblock


\bibitem[\protect\citeauthoryear{Zamani, Dumais, Craswell, Bennett, and
  Lueck}{Zamani et~al\mbox{.}}{2020a}]%
        {DBLP:conf/www/ZamaniDCBL20}
\bibfield{author}{\bibinfo{person}{Hamed Zamani}, \bibinfo{person}{Susan~T.
  Dumais}, \bibinfo{person}{Nick Craswell}, \bibinfo{person}{Paul~N. Bennett},
  {and} \bibinfo{person}{Gord Lueck}.} \bibinfo{year}{2020}\natexlab{a}.
\newblock \showarticletitle{Generating Clarifying Questions for Information
  Retrieval}. In \bibinfo{booktitle}{\emph{{WWW}}}. \bibinfo{pages}{418--428}.
\newblock


\bibitem[\protect\citeauthoryear{Zamani, Lueck, Chen, Quispe, Luu, and
  Craswell}{Zamani et~al\mbox{.}}{2020b}]%
        {DBLP:journals/corr/abs-2006-10174}
\bibfield{author}{\bibinfo{person}{Hamed Zamani}, \bibinfo{person}{Gord Lueck},
  \bibinfo{person}{Everest Chen}, \bibinfo{person}{Rodolfo Quispe},
  \bibinfo{person}{Flint Luu}, {and} \bibinfo{person}{Nick Craswell}.}
  \bibinfo{year}{2020}\natexlab{b}.
\newblock \showarticletitle{MIMICS: A Large-Scale Data Collection for Search
  Clarification}. In \bibinfo{booktitle}{\emph{{CIKM}}}.
  \bibinfo{pages}{3189–3196}.
\newblock


\bibitem[\protect\citeauthoryear{Zamani, Mitra, Chen, Lueck, Diaz, Bennett,
  Craswell, and Dumais}{Zamani et~al\mbox{.}}{2020c}]%
        {DBLP:conf/sigir/ZamaniMCLDBCD20}
\bibfield{author}{\bibinfo{person}{Hamed Zamani}, \bibinfo{person}{Bhaskar
  Mitra}, \bibinfo{person}{Everest Chen}, \bibinfo{person}{Gord Lueck},
  \bibinfo{person}{Fernando Diaz}, \bibinfo{person}{Paul~N. Bennett},
  \bibinfo{person}{Nick Craswell}, {and} \bibinfo{person}{Susan~T. Dumais}.}
  \bibinfo{year}{2020}\natexlab{c}.
\newblock \showarticletitle{Analyzing and Learning from User Interactions for
  Search Clarification}. In \bibinfo{booktitle}{\emph{{SIGIR}}}.
  \bibinfo{pages}{1181--1190}.
\newblock


\bibitem[\protect\citeauthoryear{Zhang and Balog}{Zhang and Balog}{2020}]%
        {Zhang2020}
\bibfield{author}{\bibinfo{person}{Shuo Zhang} {and} \bibinfo{person}{Krisztian
  Balog}.} \bibinfo{year}{2020}\natexlab{}.
\newblock \showarticletitle{Evaluating Conversational Recommender Systems via
  User Simulation}. In \bibinfo{booktitle}{\emph{{KDD}}}.
  \bibinfo{publisher}{{ACM}}, \bibinfo{pages}{1512--1520}.
\newblock


\bibitem[\protect\citeauthoryear{Zhang, Chen, Ai, Yang, and Croft}{Zhang
  et~al\mbox{.}}{2018}]%
        {DBLP:conf/cikm/ZhangCA0C18}
\bibfield{author}{\bibinfo{person}{Yongfeng Zhang}, \bibinfo{person}{Xu Chen},
  \bibinfo{person}{Qingyao Ai}, \bibinfo{person}{Liu Yang}, {and}
  \bibinfo{person}{W.~Bruce Croft}.} \bibinfo{year}{2018}\natexlab{}.
\newblock \showarticletitle{Towards Conversational Search and Recommendation:
  System Ask, User Respond}. In \bibinfo{booktitle}{\emph{{CIKM}}}.
  \bibinfo{pages}{177--186}.
\newblock


\end{thebibliography}

\end{document}